\documentclass[12pt]{article}
\usepackage{palatino}
\usepackage[margin=1in]{geometry}
\usepackage{graphicx}
\usepackage{amssymb}
\usepackage{amsmath}
\usepackage{epstopdf}
\usepackage{color}
\usepackage{subfiles}
\usepackage[ragged]{sidecap}
\sidecaptionvpos{figure}{t}
\usepackage{mdwlist}
\usepackage[compact]{titlesec}
\titlespacing{\part}{0pt}{0pt}{10pt}
\titlespacing{\section}{0pt}{5pt}{5pt}
\titlespacing{\subsection}{0pt}{5pt}{5pt}
\titlespacing{\subsubsection}{0pt}{*1}{*1}
\usepackage{enumitem}
\usepackage{tabulary}
\newcolumntype{K}[1]{>{\centering\arraybackslash}p{#1}}
\usepackage{titlesec}
\usepackage{breakurl}
\usepackage{wrapfig}
\usepackage{blindtext}
\titleformat{\part}{\Large\bfseries}{\thepart}{1em}{}
\usepackage{hyperref}
\usepackage{caption}
\usepackage{multicol}

\usepackage{array}
\newcolumntype{L}[1]{>{\raggedright\let\newline\\\arraybackslash\hspace{0pt}}m{#1}}
\newcolumntype{C}[1]{>{\centering\let\newline\\\arraybackslash\hspace{0pt}}m{#1}}
\newcolumntype{R}[1]{>{\raggedleft\let\newline\\\arraybackslash\hspace{0pt}}m{#1}}

\newcommand\snowmass{
\begin{center}
  \rule[-0.2in]{\hsize}{0.01in}\\
  \rule{\hsize}{0.01in}\\
  \vskip 0.1in
  Submitted to the Proceedings of the US Community Study\\ 
  on the Future of Particle Physics (Snowmass 2021)\\
  \rule{\hsize}{0.01in}\\
  \rule[+0.2in]{\hsize}{0.01in}\\[-2em]
\end{center}
}

\usepackage[firstpage=true]{background}
\backgroundsetup{contents={\parbox{6.5in}{\snowmass}}, scale=1,placement=top,opacity=1,color=black,position={3.25in,1.2in}}

\usepackage{fancyhdr}
\fancypagestyle{plain}{%
  \fancyhf{}%
  \fancyhead[C]{}
  \fancyfoot[C]{\thepage}
}

\fancypagestyle{empty}{%
  \fancyhf{}%
  \fancyhead[C]{{\it Snowmass 2021 CMB-S4 White Paper}}
  \fancyfoot[C]{\thepage}
}
\title{Snowmass 2021 CMB-S4 White Paper}
\date{}

\newcommand{\Neff}{\ensuremath{N_\mathrm{eff}}}

\definecolor{orange}{rgb}{1,0.3,0}

\newcommand{\commentout}[1]{{}}

\def\lsim{\raise-.75ex\hbox{$\buildrel<\over\sim$}}

\DeclareUrlCommand\email{\urlstyle{rm}}

\newcommand{\dsr}{{DSR}}
\newcommand{\pbdr}{{PBDR}}

\begin{document}

\maketitle

\def\bibname{References}

\bibliographystyle{utphys}  %%%% MODIFIED FOR CF %%%%

\raggedbottom

\pagenumbering{roman}

\parindent=0pt
\parskip=8pt
\setlength{\evensidemargin}{0pt}
\setlength{\oddsidemargin}{0pt}
\setlength{\marginparsep}{0.0in}
\setlength{\marginparwidth}{0.0in}
\marginparpush=0pt

\definecolor{shadecolor}{rgb}{0.8,0.90,0.95}

\pagenumbering{roman}

%\begin{center}
%Contributors and Endorsers
%\end{center}
\tolerance=4000
%%%%%%%%%%%%%%%%%%%%%%%%%%%%%%%%%%%%%%%%%%%%%%%%%%%%%%%%%%%%
%                                                          %
% Institutional aliases file.                              %
% Originally used in CV DM whitepaper,                     %
% successfully stolen for CV 21cm roadmap whitepaper,      %
% now to live a life of its own.                           %
%                                                          %
% When editing, please respect alphabetical ordering       %
% and avoid duplication (that is the entire point).        %
%                                                          %
%                                                          %
% A summer river being crossed                             %
% how pleasing                                             %
% with sandals in my hands!                                %
%       Yosa Buson (1716-1784)                             %
%                                                          %
% 08/2020: Edits to make US, UK, and Canadian addresses uniform.
%%%%%%%%%%%%%%%%%%%%%%%%%%%%%%%%%%%%%%%%%%%%%%%%%%%%%%%%%%%%

\newcommand{\Aalto}{Aalto University, FIN-00076, Aalto, Finland}
\newcommand{\Amherst}{University of Massachusetts, Amherst, MA 01003, USA}
\newcommand{\AmsterdamAstro}{Anton Pannekoek Institute for Astronomy, University of Amsterdam, 1098~XH Amsterdam, The Netherlands}
\newcommand{\ANL}{Argonne National Laboratory, Lemont, IL 60439, USA}
\newcommand{\ANLHEP}{HEP Division, Argonne National Laboratory, Lemont, IL 60439, USA}
\newcommand{\APC}{Laboratoire Astroparticule et Cosmologie (APC), CNRS/IN2P3, Universit\'e Paris Diderot, 75205 Paris Cedex 13, France}
\newcommand{\APONMSU}{Apache Point Observatory and New Mexico State University, Sunspot, NM 88349, USA}
\newcommand{\ASU}{Arizona State University, Tempe, AZ 85287, USA}
\newcommand{\AUI}{Associated Universities Incorporated, Washington, DC 20036, USA}
\newcommand{\BCCP}{Berkeley Center for Cosmological Physics, University of California, Berkeley, CA 94720, USA}
\newcommand{\Bell}{Bell Laboratories, Murray Hill, NJ 07974, USA}
\newcommand{\BenGurion}{Department of Physics, Ben-Gurion University, Be'er Sheva 84105, Israel}
\newcommand{\BNL}{Brookhaven National Laboratory, Upton, NY~11973, USA}
\newcommand{\Brown}{Brown University, Providence, RI 02912, USA}
\newcommand{\BU}{Boston University, Boston, MA 02215, USA}
\newcommand{\Buffalo}{Department of Physics, University at Buffalo, SUNY Buffalo, NY 14260 USA}
\newcommand{\Caltech}{California Institute of Technology, Pasadena, CA 91125, USA}
\newcommand{\Cantabria}{Instituto de Física de Cantabria, Cantabria, Spain}
\newcommand{\Cardiff}{School of Physics and Astronomy, Cardiff University, The Parade, Cardiff, CF24 3AA, UK}
\newcommand{\Carleton}{Carleton University, Ottawa, ON K1S 5B6, Canada}
\newcommand{\Carnegie}{The Observatories of the Carnegie Institution for Science, 813 Santa Barbara St., Pasadena, CA 91101, USA}
\newcommand{\Cavendish}{Cavendish Laboratory, University of Cambridge, Cambridge, CB3 0HE, UK}
\newcommand{\CCA}{Center for Computational Astrophysics, Flatiron Institute, New York, NY~10010, USA}
\newcommand{\CCAPP}{Center for Cosmology and AstroParticle Physics, The Ohio State University, Columbus, OH 43212}
\newcommand{\CCPP}{Center for Cosmology and Particle Physics, Department of Physics, New York University, 726 Broadway, Room 1005, New York, NY 10003, USA}
\newcommand{\CPPM}{Aix Marseille Univ, CNRS/IN2P3, CPPM, Marseille, France}
\newcommand{\CEADAP}{D\'epartement d'Astrophysique, CEA Saclay DSM/Irfu, 91191 Gif-sur-Yvette, France}
\newcommand{\CERN}{CERN, Geneva, Switzerland}
\newcommand{\CfA}{Harvard-Smithsonian Center for Astrophysics, Cambridge, MA 02138, USA}
\newcommand{\CFT}{Center for Theoretical Physics, Polish Academy of Sciences, 02-668 Warsaw, Poland}
\newcommand{\Cincinnati}{University of Cincinnati, Cincinnati, OH 45221, USA}
\newcommand{\CITA}{Canadian Institute for Theoretical Astrophysics, University of Toronto, Toronto, ON M5S 3H8, Canada}
\newcommand{\Clemson}{Clemson University, Clemson, SC 29634, USA}
\newcommand{\CNRSA}{CNRS, Laboratoire d'Annecy-le-Vieux de Physique Th\'{e}orique, Annecy-le-Vieux, France}
\newcommand{\CNYang}{C.N.\ Yang Institute for Theoretical Physics, State University of New York Stony Brook, NY~11794, USA}
\newcommand{\CNCT}{Consejo Nacional de Ciencia y Tecnolog\'ia, Av. Insurgentes Sur 1582. Colonia Cr\'edito Constructor, Del. Benito Ju\'arez, C.P. 03940, M\'exico D.F. M\'exico}
\newcommand{\CMUCosmo}{Department of Physics, McWilliams Center for Cosmology, Carnegie Mellon University, Pittsburgh, PA 15213, USA }
\newcommand{\Columbia}{Columbia University, New York, NY 10027, USA}
\newcommand{\Cornell}{Cornell University, Ithaca, NY 14853, USA}
\newcommand{\CPB}{Centre Pierre Binétruy, Berkeley, CA 94704, USA}
\newcommand{\CPthree}{CP3-Origins, 5230 Odense, Denmark}
\newcommand{\CUBoulder}{Center for Astrophysics and Space Astronomy, Department of Astrophysical and Planetary Science, University of Colorado, Boulder, CO 80309, USA}
\newcommand{\CWRU}{Case Western Reserve University, Cleveland, OH 44106, USA}
\newcommand{\daa}{David A. Dunlap Department of Astronomy and Astrophysics, University of Toronto, ON, M5S 3H4, Canada}
\newcommand{\damtp}{Department of Applied Mathematics and Theoretical Physics, University of Cambridge, Cambridge CB3 0WA, UK}
\newcommand{\Dartmouth}{Department of Physics \& Astronomy, Dartmouth College, Hanover, NH 03755, USA}
\newcommand{\DESY}{DESY, 22607 Hamburg, Germany}
\newcommand{\dfa}{Departamento de F\'{\i}sica e Astronomia, Faculdade de Ci\^{e}ncias, Universidade do Porto, Porto, Portugal}
\newcommand{\DFI}{Departamento de F\'isica, FCFM, Universidad de Chile, Blanco Encalada 2008, Santiago, Chile}
\newcommand{\DOE}{US. Department of Energy, Germantown, MD 20874, USA}
\newcommand{\drexel}{Drexel University, Philadelphia, PA 19104, USA}
\newcommand{\Duke}{Duke University and Triangle Universities Nuclear Laboratory, Durham, NC 27708, USA}
\newcommand{\DukePhys}{Department of Physics, Duke University, Durham, NC 27708, USA}
\newcommand{\dunlap}{Dunlap Institute for Astronomy and Astrophysics, University of Toronto, ON M5S 3H4, Canada}
\newcommand{\Durham}{Department of Physics, Durham University, Durham DH1 3LE, UK}
\newcommand{\EC}{Eric Chauvin Consulting Engineer}\newcommand{\ED}{University of Edinburgh, Edinburgh EH8 9YL, UK}
\newcommand{\EONS}{EONS SpA}
\newcommand{\EPFL}{Institute of Physics, Laboratory of Astrophysics, Ecole Polytechnique F\'ed\'erale de Lausanne (EPFL), Observatoire de Sauverny, 1290 Versoix, Switzerland}
\newcommand{\ESO}{European Southern Observatory, Karl-Schwarzschild-Str. 2, DE-85748 Garching b. Munchen, Germany}
\newcommand{\ETH}{ETH Zurich, Institute for Particle Physics, 8093 Zurich, Switzerland}
\newcommand{\FNAL}{Fermi National Accelerator Laboratory, Batavia, IL 60510, USA}
\newcommand{\FQAUB}{Dept. de F\' isica Qu\` antica i Astrof\' isica, Universitat de Barcelona, Mart\' i i Franqu\` es 1, E08028 Barcelona, Spain}
\newcommand{\FSU}{Florida State University, Tallahassee, FL 32306, USA}
\newcommand{\Glasgow}{University of Glasgow, Glasgow G12 8QQ, UK}
\newcommand{\GRAPPA}{GRAPPA Institute, University of Amsterdam, 1098 XH Amsterdam, The Netherlands}
\newcommand{\GSFC}{Goddard Space Flight Center, Greenbelt, MD 20771, USA}
\newcommand{\GWU}{George Washington University, Washington, DC 20052, USA}
\newcommand{\Hampton}{Hampton University, Hampton, VA 23668, USA}
\newcommand{\HarvardPhys}{Department of Physics, Harvard University, Cambridge, MA 02138, USA}
\newcommand{\Haverford}{Haverford College, Haverford, PA 19041, USA}
\newcommand{\Hawaii}{University of Hawaii, Honolulu, HI 96822, USA}
\newcommand{\HKUST}{The Hong Kong University of Science and Technology, Hong Kong SAR, China}
\newcommand{\houston}{University of Houston, Houston, TX 77204, USA}
\newcommand{\IA}{Instituto de Astrof\'{\i}sica e Ci\^encias do Espa\c{c}o (IA), Porto, Portugal}
\newcommand{\IAC}{Instituto de Astrof\'{\i}sica de Canarias, 38200 La Laguna, Tenerife, Spain}
\newcommand{\IAP}{Institut d'Astrophysique de Paris (IAP), CNRS \& Sorbonne University, Paris, France}
\newcommand{\IAS}{Institute for Advanced Study, Princeton, NJ 08540, USA}
\newcommand{\IBS}{Institute for Basic Science (IBS), Daejeon 34051, Korea}
\newcommand{\ICC}{ICC, University of Barcelona, IEEC-UB, Mart\' i i Franqu\` es, 1, E08028 Barcelona, Spain}
\newcommand{\ICCD}{Institute for Computational Cosmology, Department of Physics, Durham University, Durham DH1 3LE, UK}
\newcommand{\ICE}{Institute of Space Sciences (ICE, CSIC), Campus UAB, Carrer de Can Magrans, s/n, 08193 Barcelona, Spain}
\newcommand{\ICFUNAM}{ICFUNAM - Instituto de Ciencias F\'{i}sicas, Universidad Nacional Aut\'onoma de M\'exico,  62210 Cuernavaca, Mor., M\'exico}
\newcommand{\ICJLab}{Universit\'e Paris-Saclay, CNRS/IN2P3, IJCLab, 91405 Orsay, France}
\newcommand{\ICRR}{Institute for Cosmic Ray Resaerch, The University of Tokyo, 456 Higashi-Mozumi, Kamioka, Hida, Gifu 506-1205, Japan}
\newcommand{\ICTP}{International Centre for Theoretical Physics, Strada Costiera, 11, I-34151 Trieste, Italy}
\newcommand{\IFAE}{Institut de Fisica d'Altes Energies, The Barcelona Institute of Science and Technology, Campus UAB, 08193 Bellaterra (Barcelona), Spain}
\newcommand{\IFPU}{IFPU - Institute for Fundamental Physics of the Universe, Via Beirut 2, 34014 Trieste, Italy}
\newcommand{\IFT}{Instituto de Fisica Teorica UAM/CSIC, Universidad Autonoma de Madrid, 28049 Madrid, Spain}
\newcommand{\IFUNAM}{IFUNAM - Instituto de F\'{i}sica, Universidad Nacional Aut\'onoma de M\'exico, 04510 CDMX, M\'exico}
\newcommand{\IHEP}{Institute of High Energy Physics, Austrian Academy of Sciences, 1050 Vienna, Austria}
\newcommand{\ILL}{Institut Laue-Langevin, 71 Avenue des Martyrs, 38000 Grenoble, France}
\newcommand{\Imperial}{Theoretical Physics, Blackett Laboratory, Imperial College, London SW7 2AZ, UK}
\newcommand{\Indiana}{Indiana University, Bloomington, IN 47405, USA}
\newcommand{\INAFOATs}{INAF - Osservatorio Astronomico di Trieste, Via G.B. Tiepolo 11, 34143 Trieste, Italy}
\newcommand{\INAFOAS}{INAF - Osservatorio di Astrofisica e Scienza dello Spazio di Bologna, via Piero Gobetti 93/3, I-40129 Bologna, Italy}
\newcommand{\INFNCag}{Istituto Nazionale di Fisica Nucleare, Sezione di Cagliari,  09126 Cagliari, Italy}
\newcommand{\INFNCat}{Istituto Nazionale di Fisica Nucleare, Sezione di Catania, 95125 Catania, Italy}
\newcommand{\INFNG}{Istituto Nazionale di Fisica Nucleare, Sezione di Genova, 16146 Genova, Italy}
\newcommand{\INFN}{INFN - National Institute for Nuclear Physics, Via Valerio 2, I-34127 Trieste, Italy}
\newcommand{\INFNFE}{Istituto Nazionale di Fisica Nucleare, Sezione di Ferrara, 40122, Italy }
\newcommand{\INFNLNF}{Istituto Nazionale di Fisica Nucleare, Laboratori Nazionali di Frascati, 00044 Frascati, Italy}
\newcommand{\INFNLNS}{Istituto Nazionale di Fisica Nucleare, Laboratori Nazionali del Sud, 95125 Catania, Italy}
\newcommand{\INFNN}{Istituto Nazionale di Fisica Nucleare, Sezione di Napoli, 80125 Napoli, Italy }
\newcommand{\INFNPD}{Istituto Nazionale di Fisica Nucleare, Sezione di Padova, 35131 Padova, Italy}
\newcommand{\INFNRM}{Istituto Nazionale di Fisica Nucleare, Sezione di Roma, 00185 Roma, Italy}
\newcommand{\INFNT}{Istituto Nazionale di Fisica Nucleare, Sezione di Torino, 10125, Italy }
\newcommand{\IoA}{Institute of Astronomy, University of Cambridge, Cambridge CB3 0HA, UK}
\newcommand{\IPP}{Institute for Particle Physics, Victoria, BC V8W 3P6, Canada}
\newcommand{\IPMU}{Kavli Institute for the Physics and Mathematics of the Universe, University of Tokyo, Kashiwa, Japan}
\newcommand{\IPNL}{Universit\'e de Lyon, F-69622, Lyon, France; Universit\'e de Lyon 1, Villeurbanne; CNRS/IN2P3, Institut de Physique Nucl\'eaire de Lyon}
\newcommand{\IRFU}{IRFU, CEA, Universit\'e Paris-Saclay, F-91191 Gif-sur-Yvette, France}
\newcommand{\ITFA}{Institute for Theoretical Physics, University of Amsterdam, Science Park 904, 1098 XH Amsterdam, The Netherlands}
\newcommand{\IUCAA}{The Inter-University Centre for Astronomy and Astrophysics, Pune, 411007, India}
\newcommand{\JAXA}{ISAS JAXA, Sagamihara, Kanagawa 252-5210, Japan}
\newcommand{\Jerusalem}{Hebrew University of Jerusalem, 91904 Jerusalem, Israel}
\newcommand{\JHU}{Johns Hopkins University, Baltimore, MD 21218, USA}
\newcommand{\JLAB}{Thomas Jefferson National Laboratory, Newport News, VA 23606, USA}
\newcommand{\JPL}{Jet Propulsion Laboratory, California Institute of Technology, Pasadena, CA 91011, USA}
\newcommand{\KASSI}{Korea Astronomy and Space Science Institute, Daejeon 34055, Korea}
\newcommand{\kavli}{Kavli Institute for Cosmology, University of Cambridge, Cambridge CB3 0HA, UK}
\newcommand{\KEK}{High Energy Accelerator Research Organization (KEK), Tsukuba, Ibaraki 305-0801, Japan}
\newcommand{\KIAS}{School of Physics, Korea Institute for Advanced Study, Dongdaemun-gu, Seoul 130-722, Korea}
\newcommand{\KICP}{Kavli Institute for Cosmological Physics, University of Chicago, Chicago, IL 60637, USA}
\newcommand{\KIPAC}{Kavli Institute for Particle Astrophysics and Cosmology, Stanford, CA 94305, USA}
\newcommand{\KINGS}{King's College London, London WC2R 2LS, UK}
\newcommand{\Kobe}{Kobe University, 657-8501 Kobe, Japan}
\newcommand{\KPH}{Johannes Gutenberg University, 55128 Mainz, Germany}
\newcommand{\KPMU}{University of Tokyo, 277-8583  Kashiwa, Japan}
\newcommand{\KSU}{Kansas State University, Manhattan, KS 66506, USA}
\newcommand{\KwaZuluNatal}{Astrophysics and Cosmology Research Unit, School of Chemistry and Physics, University of KwaZulu-Natal, Durban 4000, South Africa}
\newcommand{\Kyoto}{Department of Physics, Kyoto University, Kyoto 606-8502, Japan}
\newcommand{\Lafayette}{Lafayette College, Easton, PA 18042, USA}
\newcommand{\LANL}{Los Alamos National Laboratory, Los Alamos, NM 87545, USA}
\newcommand{\LBL}{Lawrence Berkeley National Laboratory, Berkeley, CA 94720, USA}
\newcommand{\Leiden}{Lorentz Institute, Leiden University, Niels Bohrweg 2,Leiden, NL 2333 CA, The Netherlands}
\newcommand{\Liverpool}{University of Liverpool, Liverpool L69 7ZE, UK}
\newcommand{\LJMU}{Liverpool John Moores University, Liverpool L3 5RF, UK}
\newcommand{\LLNL}{Lawrence Livermore National Laboratory, Livermore, CA, 94550, USA}
\newcommand{\LMU}{Ludwig-Maximilians-Universit\"at, 81679 Munich, Germany}
\newcommand{\LPC}{Universit\'e Clermont Auvergne, CNRS/IN2P3, Laboratoire de Physique de Clermont, F-63000 Clermont-Ferrand, France}
\newcommand{\NPNHE}{Sorbonne Universit\'e, Universit\'e Paris Diderot, CNRS/IN2P3, Laboratoire de Physique Nucl\'eaire et de Hautes Energies, LPNHE, F-75252 Paris, France}
\newcommand{\McGill}{McGill University, Montreal, QC H3A 2T8, Canada}
\newcommand{\Melbourne}{School of Physics, The University of Melbourne, Parkville, VIC 3010, Australia}
\newcommand{\MSU}{Michigan State University, East Lansing, MI 48824, USA}
\newcommand{\Miller}{Miller Institute for Basic Research in Science, 468 Donner Lab, Berkeley, CA 94720, USA}
\newcommand{\Mines}{Colorado School of Mines, Golden, CO 80401, USA}
\newcommand{\MIT}{Massachusetts Institute of Technology, Cambridge, MA 02139, USA}
\newcommand{\MilanoBicocca}{Department of Physics, University of Milano - Bicocca, Piazza della Scienza 3, I-20126 Milano, Italy}
\newcommand{\MPE}{Max-Planck-Institut f\"{u}r extraterrestrische Physik (MPE), Giessenbachstrasse 1, D-85748 Garching bei M\"unchen, Germany}
\newcommand{\MPIA}{Max-Planck-Institut f\"{u}r Astrophysik, Karl-Schwarzschild-Str. 1, 85741 Garching, Germany}
\newcommand{\MPP}{Max-Planck-Institut f\"{u}r Physik (Werner-Heisenberg-Institut), F\"ohringer Ring 6, D-80805 M\"unchen, Germany}
\newcommand{\LAM}{Aix Marseille Univ, CNRS, CNES, LAM, Marseille, France}
\newcommand{\LUPM}{Laboratoire Univers et Particules de Montpellier, Univ. Montpellier and CNRS, 34090 Montpellier, France}
\newcommand{\NAOC}{National Astronomical Observatories, Chinese Academy of Sciences, PR China}
\newcommand{\NCBJ}{National Center for Nuclear Research, Ul.Pasteura 7,Warsaw, Poland}
\newcommand{\NCU}{National Central University, Taoyuan City 32001, Taiwan (R.O.C.)}
\newcommand{\NCSU}{Physics Department, North Carolina State University, Raleigh, NC 27607, USA}
\newcommand{\ND}{University of Notre Dame, Notre Dame, IN 46556, USA}
\newcommand{\NIST}{National Institute of Standards and Technology, Boulder, CO 80305, USA}
\newcommand{\NIU}{Northern Illinois University, DeKalb, Illinois 60115, USA}
\newcommand{\NMSU}{New Mexico State University, Las Cruces, NM 88003, USA}
\newcommand{\NOAO}{National Optical Astronomy Observatory, Tucson, AZ 85719 USA}
\newcommand{\Northwestern}{Northwestern University, Evanston, IL 60201, USA}
\newcommand{\Nottingham}{University of Nottingham, Nottingham NG7 2RD, UK}
\newcommand{\NPPSFAmes}{NASA Postdoctoral Program Senior Fellow, NASA Ames Research Center, Moffett Field, CA 94035, USA}
\newcommand{\NRC}{Herzberg Astronomuy and Astrophyiscs - NRC, Victoria, Canada} 
\newcommand{\NWU}{Northwestern University, Evanston, IL 60208, USA}
\newcommand{\NYU}{New York University, New York, NY 10003, USA}
\newcommand{\OIRLab}{NSF's National Optical-Infrared Astronomy Research Laboratory, Tucson, AZ 85719, USA}
\newcommand{\OK}{ University of Oklahoma, Norman, OK 73019, USA}
\newcommand{\ORNL}{Oak Ridge National Laboratory, Oak Ridge, TN 37831, USA}
\newcommand{\OSU}{The Ohio State University, Columbus, OH 43212, USA}
\newcommand{\OU}{Department of Physics and Astronomy, Ohio University, Clippinger Labs, Athens, OH 45701, USA}
\newcommand{\OskarKlein}{Oskar Klein Centre for Cosmoparticle Physics, Stockholm University, AlbaNova, Stockholm SE-106 91, Sweden}
\newcommand{\Oslo}{University of Oslo, Oslo, WPQC XM, Norway}
\newcommand{\Oxford}{University of Oxford, Oxford OX1~3RH, UK}
\newcommand{\Oxy}{Occidental College, Los Angeles, CA 90041, USA}
\newcommand{\ParisSud}{Universit\'{e} Paris-Sud, LAL, UMR 8607, F-91898 Orsay Cedex, France \& CNRS/IN2P3, F-91405 Orsay, France}
\newcommand{\PI}{Perimeter Institute, Waterloo, ON N2L 2Y5, Canada}
\newcommand{\Pitt}{University of Pittsburgh and PITT PACC, Pittsburgh, PA 15260, USA}
\newcommand{\PNNL}{Pacific Northwest National Laboratory, Richland, WA 99352, USA}
\newcommand{\PNPI}{Petersburg Nuclear Physics Institute, 188300 Gatchina, Russia}
\newcommand{\Port}{Institute of Cosmology \& Gravitation, University of Portsmouth, Portsmouth PO1 3FX, UK}
\newcommand{\Princeton}{Princeton University, Princeton, NJ 08544, USA}
\newcommand{\PSU}{The Pennsylvania State University, University Park, PA 16802, USA}
\newcommand{\PSUAstro}{Department of Astronomy and Astrophysics, The Pennsylvania State University, University Park, PA 16802, USA}
\newcommand{\PSUIGC}{Institute for Gravitation and the Cosmos, The Pennsylvania State University, University Park, PA 16802, USA}
\newcommand{\PU}{Pontificia Universidad Católica de Chile, Santiago, Región Metropolitana, Chile}
\newcommand{\Purdue}{Purdue University, West Lafayette, IN 47907, USA}
\newcommand{\PW}{Participation Worldscope, Sedona, Arizona and Hong Kong, SAR PRC}
\newcommand{\Queens}{Queen's University, Kingston, ON K7L 3N6, Canada}
\newcommand{\Queensland}{The University of Queensland, School of Mathematics and Physics, QLD 4072, Australia}
\newcommand{\QMUL}{Queen Mary University of London, London E1~4NS, UK}
\newcommand{\RAL}{Radio Astronomy Laboratory, University of California Berkeley, Berkeley, CA 94720, USA}
\newcommand{\Rice}{Department of Physics \& Astronomy, Rice University, Houston, Texas 77005, USA}
\newcommand{\RSS}{Remote Science Services, LLC}
\newcommand{\RIT}{Rochester Institute of Technology, Rochester, NY 14623, USA}
\newcommand{\RomaS}{Dipartimento di Fisica, Universit\`{a} La Sapienza, P. le A. Moro 2, Roma, Italy}
\newcommand{\Rome}{University of Rome Tor Vergata, 00133 Roma RM, Italy}
\newcommand{\RRI}{Raman Research Institute, Bengaluru, Karnataka 560080, India}
\newcommand{\RUG}{Kapteyn Astronomical Institute, University of Groningen, 9700 AV Groningen, The Netherlands}
\newcommand{\Rutgers}{Department of Physics and Astronomy, Rutgers, the State University of New Jersey, Piscataway, NJ 08854, USA}
\newcommand{\SAAO}{South African Astronomical Observatory, PO Box 9, Observatory 7935, Cape Town, South Africa}
\newcommand{\SAO}{Smithsonian Astrophysical Observatory, Cambridge, MA 02138, USA}
\newcommand{\SAIMSU}{Sternberg Astronomical Institute, Moscow State University, Moscow, 119992, Russia}
\newcommand{\Sanford}{Sanford Underground Research Facility, Lead, SD 57754, USA}
\newcommand{\Sassari}{Universit\`a di Sassari, 07100 Sassari,  Italy}
\newcommand{\SCIPP}{University of California, Santa Cruz, Santa Cruz, CA 95064, USA}
\newcommand{\Sejong}{Department of Physics and Astronomy, Sejong University, Seoul, 143-747, Korea}
\newcommand{\Sheffield}{University of Sheffield, Sheffield S3 7RH, UK}
\newcommand{\SHAO}{Shanghai Astronomical Observatory (SHAO), Nandan Road 80, Shanghai 200030, China}
\newcommand{\Siena}{Siena College, Loudonville, NY 12211, USA}
\newcommand{\SISSA}{SISSA - International School for Advanced Studies, Via Bonomea 265, 34136 Trieste, Italy}
\newcommand{\SimonFraser}{Department of Physics, Simon Fraser University, Burnaby, BC  V5A 1S6, Canada}
\newcommand{\SLAC}{SLAC National Accelerator Laboratory, Menlo Park, CA 94025, USA}
\newcommand{\SMU}{Southern Methodist University, Dallas, TX 75275, USA}
\newcommand{\SNOLAB}{SNOLAB, Lively, ON P3Y 1N2, Canada}
\newcommand{\SoCal}{University of Southern California, CA 90089, USA}
\newcommand{\StAndrews}{School of Physics and Astronomy, University of St Andrews, North Haugh, St Andrews KY16 9SS, UK}
\newcommand{\Stanford}{Stanford University, Stanford, CA 94305, USA}
\newcommand{\StonyBrook}{Stony Brook University, Stony Brook, NY 11794, USA}
\newcommand{\STSCI}{Space Telescope Science Institute, Baltimore, MD 21218, USA}
\newcommand{\SUNYA}{University at Albany SUNY, Albany, NY 12222, USA}
\newcommand{\SUOT}{Swinburne University of Technology, Centre for Astrophysics and Supercomputing, Melbourne, VIC 3122, Australia}
\newcommand{\SussexAstronomy}{Astronomy Centre, School of Mathematical and Physical Sciences, University of Sussex, Brighton BN1 9QH, UK}
\newcommand{\Syracuse}{Syracuse University, Syracuse, NY 13244, USA}
\newcommand{\Tamu}{Texas A\&M University, College Station, TX 77843, USA }
\newcommand{\TCU}{Department of Physics \& Astronomy, Texas Christian University, Fort Worth, TX 76129, USA}
\newcommand{\Techsource}{Techsource Incorporated, Los Alamos, NM 87544, USA}
\newcommand{\TelAviv}{Tel-Aviv University,  69978 Tel-Aviv, Israel}
\newcommand{\Temple}{Temple University, Philadelphia, PA 19122, USA}
\newcommand{\TIFR}{Tata Institute of Fundamental Research, Homi Bhabha Road, Mumbai 400005 India}
\newcommand{\Tsinghua}{Department of Physics and Tsinghua Center for Astrophysics, Tsinghua University, Beijing 100084, P R China}
\newcommand{\TTU}{Texas Tech University, Lubbock, TX 79409, USA}
\newcommand{\TUM}{Technical University of Munich, 80333 Munich, Germany}
\newcommand{\Turku}{University of Turku, Turku, Finland}
\newcommand{\UA}{University of Alabama, Tuscaloosa, AL 35487, USA}
\newcommand{\UAS}{Department of Astronomy/Steward Observatory, University of Arizona, Tucson, AZ  85721, USA}
\newcommand{\UAM}{Universidad Aut\'onoma de Madrid, 28049, Madrid, Spain}
\newcommand{\UBC}{University of British Columbia, Vancouver, BC V6T 1Z1, Canada}
\newcommand{\UCB}{University of California, Berkeley, CA 94720, USA}
\newcommand{\UCBA}{Department of Astronomy, University of California, Berkeley, CA 94720, USA}
\newcommand{\UCBP}{Department of Physics, University of California, Berkeley, CA 94720, USA}
\newcommand{\UCBSSL}{Space Sciences Laboratory, University of California, Berkeley, CA 94720, USA}
\newcommand{\UCD}{University of California, Davis, CA 95616, USA}
\newcommand{\UChicago}{University of Chicago, Chicago, IL 60637, USA}
\newcommand{\UCI}{University of California, Irvine, CA 92697, USA}
\newcommand{\UCLA}{University of California, Los Angeles, Los Angeles, CA 90095, USA}
\newcommand{\UCL}{University College London, London WC1E 6BT, UK}
\newcommand{\UCR}{University of California, Riverside, CA 92521, USA}
\newcommand{\UCSB}{University of California, Santa Barbara, CA 93106, USA}
\newcommand{\UCSC}{University of California, Santa Cruz, CA 95064, USA}
\newcommand{\UCSD}{University of California, San Diego, La Jolla, CA 92093, USA}
\newcommand{\UCT}{Department of Astronomy, University of Cape Town, Rondebosch 7701, South Africa}
\newcommand{\UFerrara}{Dipartimento di Fisica e Scienze della Terra, Universit\`a di Ferrara, Polo Scientifico e Tecnologico - Edificio C Via Saragat 1, 44122 Ferrara, Italy}
\newcommand{\UFL}{University of Florida, Gainesville, FL 32611, USA}
\newcommand{\Unige}{D\'epartement de Physique Th\'eorique et CAP, Universit\'e de Gen\`eve, CH-1211 Gen\`eve 4, Switzerland}
\newcommand{\UFN}{Universit\`a Federico II di Napoli, 80125 Napoli, Italy}
\newcommand{\UGTO}{Divisi\'on de Ciencias e Ingenier\'ias, Universidad de Guanajuato, Le\'on 37150, M\'exico}
\newcommand{\UKY}{University of Kentucky, Lexington, KY 40506, USA}
\newcommand{\UMD}{University of Maryland, College Park, MD 20742, USA}
\newcommand{\UMiami}{University of Miami, Coral Gables, FL 33124, USA}
\newcommand{\UMich}{University of Michigan, Ann Arbor, MI 48109, USA}
\newcommand{\UMN}{University of Minnesota, Minneapolis, MN 55455, USA}
\newcommand{\UnB}{Instituto de F\'{i}sica, Universidade de Bras\'{i}lia, 70919-970, Bras\'{i}lia, DF, Brazil}
\newcommand{\UNC}{University of North Carolina at Chapel Hill, Chapel Hill, NC 27599, USA}
\newcommand{\UNH}{University of New Hampshire, Durham, NH 03824, USA}
\newcommand{\UNIMI}{Dipartimento di Fisica ``Aldo Pontremoli'', Universit\`a{} degli Studi di Milano, via Celoria 16, 20133 Milano, Italy}
\newcommand{\UNIPD}{Dipartimento di Fisica e Astronomia ``G. Galilei'', Universit\`a degli Studi di Padova, 35131~Padova, Italy}
\newcommand{\UNM}{Department of Physics and Astronomy, University of New Mexico, Albuquerque, NM 87131, USA}
\newcommand{\UNV}{University of Nevada, Reno, NV 89557, USA}
\newcommand{\UoM}{Jodrell Bank Center for Astrophysics, School of Physics and Astronomy, University of Manchester, Manchester M13 9PL, UK}
\newcommand{\UPenn}{Department of Physics and Astronomy, University of Pennsylvania, Philadelphia, PA 19104, USA}
\newcommand{\UR}{Department of Physics and Astronomy, University of Rochester, Rochester, NY 14627, USA}
\newcommand{\UrbanaC}{Illinois Center for Advanced Studies of the Universe \& Department of Physics, University of Illinois at Urbana-Champaign, Urbana, IL 61801, USA}
\newcommand{\USC}{The University of South Carolina, Columbia, SC 29208, USA}
\newcommand{\USD}{The University of South Dakota, Vermillion, SD 57069, USA}
\newcommand{\UTD}{University of Texas at Dallas, Texas 75080, USA}
\newcommand{\UT}{University of Texas at Austin, Austin, TX 78712, USA}
\newcommand{\UTAustinth}{Theory Group, Department of Physics, University of Texas at Austin, Austin, TX 78712, USA}
\newcommand{\Utenn}{The University of Tennessee, Knoxville, TN 37996, USA}
\newcommand{\Utah}{ Department of Physics and Astronomy, University of Utah, Salt Lake City, UT 84112, USA}
\newcommand{\UVA}{University of Virginia, Charlottesville, VA 22903, USA}
\newcommand{\UVic}{University of Victoria, Victoria, BC V8P 5C2, Canada}
\newcommand{\UWaterloo}{Department of Physics and Astronomy, University of Waterloo, Waterloo, ON N2L 3G1, Canada}
\newcommand{\UWisc}{Department of Astronomy, University of Wisconsin-Madison, Madison, WI 53706, USA}
\newcommand{\UWMadison}{Department of Physics, University of Wisconsin-Madison, Madison, WI 53706, USA}
\newcommand{\UW}{University of Washington, Seattle, WA 98195, USA}
\newcommand{\UWC}{Department of Physics \& Astronomy, University of the Western Cape, Cape Town 7535, South Africa}
\newcommand{\USTCH}{Department of Astronomy, School of Physical Sciences, University of Science and Technology of China, Hefei, Anhui 230026, China}
\newcommand{\Vanderbilt}{Physics \& Astronomy Department, Vanderbilt University, Nashville, TN 37235, USA}
\newcommand{\VSI}{Van Swinderen Institute for Particle Physics and Gravity, University of Groningen, 9747~AG~Groningen, The~Netherlands}
\newcommand{\VT}{Virginia Tech, Blacksburg, VA 24061, USA}
\newcommand{\VUU}{Virginia Union University, Richmond, Virginia, 23220, USA}
\newcommand{\WCA}{Waterloo Centre for Astrophysics, University of Waterloo, Waterloo, ON N2L 3G1, Canada} 
\newcommand{\Weizmann}{Weizmann Institute of Science, 76100 Rehovot, Israel}
\newcommand{\Wellesley}{Wellesley College, Wellesley, MA 02481, USA}
\newcommand{\wiscIce}{University of Wisconsin, Madison, WI 53706, USA}
\newcommand{\WM}{College of William and Mary, Newport News, VA 23606, USA}
\newcommand{\WUSL}{Washington University in St Louis, St. Louis, MO 63130, USA}
\newcommand{\McDoWUSL}{McDonnell Center for the Space Sciences, Washington University, St. Louis, MO 63130, USA}
\newcommand{\WVU}{CSEE, West Virginia University, Morgantown, WV 26505, USA}
\newcommand{\WVUGWAC}{Center for Gravitational Waves and Cosmology, West Virginia University, Morgantown, WV 26505, USA}
\newcommand{\Wyoming}{Department of Physics and Astronomy, University of Wyoming, Laramie, WY 82071, USA}
\newcommand{\Yale}{Department of Physics, Yale University, New Haven, CT 06520, USA}
\newcommand{\YorkU}{Department of Physics and Astronomy, York University, Toronto, ON M3J 1P3, Canada}

% This file is automatically generated by a script, so edits may be overwritten. 
% author.py --optin ../2022_full_list/CMB-S4_Membership_Members_20220313_851am.csv --rulesfile CMB-S4_rules.pkl

\newcounter{CMBSFourAffiliationCount}
\newcommand{\CMBSFourAffiliation}[1]{\refstepcounter{CMBSFourAffiliationCount}\label{#1}}
\CMBSFourAffiliation{UCIrvine}
\CMBSFourAffiliation{UniversityofMelbourne}
\CMBSFourAffiliation{JohnsHopkinsUniversity}
\CMBSFourAffiliation{UniversityofIllinoisatUrbana-Champaign}
\CMBSFourAffiliation{SLAC}
\CMBSFourAffiliation{ClemsonUniversity}
\CMBSFourAffiliation{StanfordUniversity}
\CMBSFourAffiliation{OxfordUniversity}
\CMBSFourAffiliation{UCBerkeley}
\CMBSFourAffiliation{LawrenceBerkeleyNationalLaboratory}
\CMBSFourAffiliation{RiceUniversity}
\CMBSFourAffiliation{UniversityofBritishColumbia}
\CMBSFourAffiliation{Fermilab}
\CMBSFourAffiliation{UCSanDiego}
\CMBSFourAffiliation{CenterforAstrophysics|HarvardandSmithsonian}
\CMBSFourAffiliation{YaleUniversity}
\CMBSFourAffiliation{SISSA}
\CMBSFourAffiliation{IFPU}
\CMBSFourAffiliation{HarvardUniversity}
\CMBSFourAffiliation{UniversityofNewMexico}
\CMBSFourAffiliation{ArgonneNationalLaboratory}
\CMBSFourAffiliation{CardiffUniversity}
\CMBSFourAffiliation{AstroParticleandCosmologyLaboratory}
\CMBSFourAffiliation{Caltech}
\CMBSFourAffiliation{CornellUniversity}
\CMBSFourAffiliation{NIST}
\CMBSFourAffiliation{UniversityofGeneva}
\CMBSFourAffiliation{InstitutdAstrophysiquedeParis}
\CMBSFourAffiliation{KICP}
\CMBSFourAffiliation{UniversityofChicago}
\CMBSFourAffiliation{LAMMarseille}
\CMBSFourAffiliation{CentrePierreBinétruy}
\CMBSFourAffiliation{UniversityofCincinnati}
\CMBSFourAffiliation{AssociatedUniversitiesIncorporated}
\CMBSFourAffiliation{JPL}
\CMBSFourAffiliation{LMUMunich}
\CMBSFourAffiliation{UniversityofTexasatAustin}
\CMBSFourAffiliation{CITA}
\CMBSFourAffiliation{UniversitadiFerrara}
\CMBSFourAffiliation{INFN}
\CMBSFourAffiliation{UniversityofManchester}
\CMBSFourAffiliation{ArizonaStateUniversity}
\CMBSFourAffiliation{DartmouthCollege}
\CMBSFourAffiliation{UniversityofCambridge}
\CMBSFourAffiliation{UniversityofToronto}
\CMBSFourAffiliation{EricChauvinConsultingEngineer}
\CMBSFourAffiliation{UniversityofKwazulu-Natal}
\CMBSFourAffiliation{UniversityofTokyo}
\CMBSFourAffiliation{KIPAC}
\CMBSFourAffiliation{UniversityofMilano-Bicocca}
\CMBSFourAffiliation{FlatironInstitute}
\CMBSFourAffiliation{KEK}
\CMBSFourAffiliation{UniversityofPennsylvania}
\CMBSFourAffiliation{UniversityofSheffield}
\CMBSFourAffiliation{PrincetonUniversity}
\CMBSFourAffiliation{PontificiaUniversidadCatólicadeChile}
\CMBSFourAffiliation{NorthwesternUniversity}
\CMBSFourAffiliation{UniversityofArizona}
\CMBSFourAffiliation{NASAGoddardSpaceFlightCenter}
\CMBSFourAffiliation{SimonFraserUniversity}
\CMBSFourAffiliation{UniversityofSouthernCalifornia}
\CMBSFourAffiliation{HaverfordCollege}
\CMBSFourAffiliation{NorthCarolinaStateUniversity}
\CMBSFourAffiliation{StockholmUniversity}
\CMBSFourAffiliation{CaseWesternReserveUniversity}
\CMBSFourAffiliation{UniversityofColoradoBoulder}
\CMBSFourAffiliation{UniversityofMinnesota}
\CMBSFourAffiliation{FloridaStateUniversity}
\CMBSFourAffiliation{Columbia}
\CMBSFourAffiliation{Cavendish}
\CMBSFourAffiliation{SorbonneUniversité}
\CMBSFourAffiliation{DunlapInstitute}
\CMBSFourAffiliation{MillerInstitute}
\CMBSFourAffiliation{UniversityofVirginia}
\CMBSFourAffiliation{HerzbergAstronomyandAstrophysicsNRC}
\CMBSFourAffiliation{UniversityofVictoria}
\CMBSFourAffiliation{UCSantaBarbara}
\CMBSFourAffiliation{KavliIPMU}
\CMBSFourAffiliation{WashingtonUniversitySt.Louis}
\CMBSFourAffiliation{UCDavis}
\CMBSFourAffiliation{UniversityofPittsburgh}
\CMBSFourAffiliation{BenGurionUniversity}
\CMBSFourAffiliation{AaltoUniversity}
\CMBSFourAffiliation{UniversityofSussex}
\CMBSFourAffiliation{UniversityofTurku}
\CMBSFourAffiliation{IJCLab}
\CMBSFourAffiliation{UniversityofWashington}
\CMBSFourAffiliation{TexasTechUniversity}
\CMBSFourAffiliation{PerimeterInstitute}
\CMBSFourAffiliation{ISASJAXA}
\CMBSFourAffiliation{UniversityofGroningen}
\CMBSFourAffiliation{CEASaclay}
\CMBSFourAffiliation{SouthernMethodistUniversity}
\CMBSFourAffiliation{StonyBrookUniversity}
\CMBSFourAffiliation{EuropeanSouthernObservatory}
\CMBSFourAffiliation{UniversityofAmsterdam}
\CMBSFourAffiliation{UniversityofWisconsinMadison}
\CMBSFourAffiliation{RSS}
\CMBSFourAffiliation{SpaceTelescopeScienceInstitute}
\CMBSFourAffiliation{EONSSpA}
\CMBSFourAffiliation{UniversityofRomeTorVergata}
\CMBSFourAffiliation{Aix-MarseilleUniversity}
\CMBSFourAffiliation{InstitutodeFísicadeCantabria}
\CMBSFourAffiliation{BrookhavenNationalLaboratory}
\CMBSFourAffiliation{RamanResearchInstitute}
\CMBSFourAffiliation{PennsylvaniaStateUniversity}
\CMBSFourAffiliation{KyotoUniversity}
\CMBSFourAffiliation{UniversitadegliStudidiMilan}
\CMBSFourAffiliation{BrownUniversity}
\CMBSFourAffiliation{InstituteforAdvancedStudy}
\CMBSFourAffiliation{SyracuseUniversity}
\CMBSFourAffiliation{UniversityofOslo}
\CMBSFourAffiliation{MichiganStateUniversity}
\CMBSFourAffiliation{SmithsonianAstrophysicalObservatory}
\CMBSFourAffiliation{BellLaboratories}
\CMBSFourAffiliation{MIT}
The CMB-S4 Collaboration:
Kevork Abazajian,\textsuperscript{\ref{UCIrvine}}
Arwa Abdulghafour,\textsuperscript{\ref{UniversityofMelbourne}}
Graeme~E.\ Addison,\textsuperscript{\ref{JohnsHopkinsUniversity}}
Peter Adshead,\textsuperscript{\ref{UniversityofIllinoisatUrbana-Champaign}}
Zeeshan Ahmed,\textsuperscript{\ref{SLAC}}
Marco Ajello,\textsuperscript{\ref{ClemsonUniversity}}
Daniel Akerib,\textsuperscript{\ref{SLAC}}
Steven~W.\ Allen,\textsuperscript{\ref{StanfordUniversity},\ref{SLAC}}
David Alonso,\textsuperscript{\ref{OxfordUniversity}}
Marcelo Alvarez,\textsuperscript{\ref{UCBerkeley},\ref{LawrenceBerkeleyNationalLaboratory}}
Mustafa A.~Amin,\textsuperscript{\ref{RiceUniversity}}
Mandana Amiri,\textsuperscript{\ref{UniversityofBritishColumbia}}
Adam Anderson,\textsuperscript{\ref{Fermilab}}
Behzad Ansarinejad,\textsuperscript{\ref{UniversityofMelbourne}}
Melanie Archipley,\textsuperscript{\ref{UniversityofIllinoisatUrbana-Champaign}}
Kam~S.\ Arnold,\textsuperscript{\ref{UCSanDiego}}
Matt Ashby,\textsuperscript{\ref{CenterforAstrophysics|HarvardandSmithsonian}}
Han Aung,\textsuperscript{\ref{YaleUniversity}}
Carlo Baccigalupi,\textsuperscript{\ref{SISSA},\ref{IFPU}}
Carina Baker,\textsuperscript{\ref{UniversityofIllinoisatUrbana-Champaign}}
Abhishek Bakshi,\textsuperscript{\ref{Fermilab}}
Debbie Bard,\textsuperscript{\ref{LawrenceBerkeleyNationalLaboratory}}
Denis Barkats,\textsuperscript{\ref{CenterforAstrophysics|HarvardandSmithsonian},\ref{HarvardUniversity}}
Darcy Barron,\textsuperscript{\ref{UniversityofNewMexico}}
Peter S.~Barry,\textsuperscript{\ref{ArgonneNationalLaboratory},\ref{CardiffUniversity}}
James~G.\ Bartlett,\textsuperscript{\ref{AstroParticleandCosmologyLaboratory}}
Paul Barton,\textsuperscript{\ref{LawrenceBerkeleyNationalLaboratory}}
Ritoban Basu Thakur,\textsuperscript{\ref{Caltech}}
Nicholas Battaglia,\textsuperscript{\ref{CornellUniversity}}
Jim Beall,\textsuperscript{\ref{NIST}}
Rachel Bean,\textsuperscript{\ref{CornellUniversity}}
Dominic Beck,\textsuperscript{\ref{StanfordUniversity}}
Sebastian Belkner,\textsuperscript{\ref{UniversityofGeneva}}
Karim  Benabed,\textsuperscript{\ref{InstitutdAstrophysiquedeParis}}
Amy~N.\ Bender,\textsuperscript{\ref{ArgonneNationalLaboratory},\ref{KICP}}
Bradford~A.\ Benson,\textsuperscript{\ref{Fermilab},\ref{UniversityofChicago}}
Bobby Besuner,\textsuperscript{\ref{LawrenceBerkeleyNationalLaboratory}}
Matthieu Bethermin,\textsuperscript{\ref{LAMMarseille}}
Sanah Bhimani,\textsuperscript{\ref{YaleUniversity}}
Federico Bianchini,\textsuperscript{\ref{StanfordUniversity},\ref{SLAC}}
Simon Biquard,\textsuperscript{\ref{AstroParticleandCosmologyLaboratory},\ref{CentrePierreBinétruy}}
Ian Birdwell,\textsuperscript{\ref{UniversityofNewMexico}}
Colin~A.\ Bischoff,\textsuperscript{\ref{UniversityofCincinnati}}
Lindsey Bleem,\textsuperscript{\ref{ArgonneNationalLaboratory},\ref{KICP}}
Paulina Bocaz,\textsuperscript{\ref{AssociatedUniversitiesIncorporated}}
James~J.\ Bock,\textsuperscript{\ref{Caltech},\ref{JPL}}
Sebastian Bocquet,\textsuperscript{\ref{LMUMunich}}
Kimberly K.~Boddy,\textsuperscript{\ref{UniversityofTexasatAustin}}
J.~Richard Bond,\textsuperscript{\ref{CITA}}
Julian Borrill,\textsuperscript{\ref{LawrenceBerkeleyNationalLaboratory},\ref{UCBerkeley}}
Fran\c{c}ois~R.\ Bouchet,\textsuperscript{\ref{InstitutdAstrophysiquedeParis}}
Thejs Brinckmann,\textsuperscript{\ref{UniversitadiFerrara},\ref{INFN}}
Michael~L.\ Brown,\textsuperscript{\ref{UniversityofManchester}}
Sean Bryan,\textsuperscript{\ref{ArizonaStateUniversity}}
Victor Buza,\textsuperscript{\ref{UniversityofChicago},\ref{KICP}}
Karen Byrum,\textsuperscript{\ref{ArgonneNationalLaboratory}}
Erminia Calabrese,\textsuperscript{\ref{CardiffUniversity}}
Victoria Calafut,\textsuperscript{\ref{CITA}}
Robert Caldwell,\textsuperscript{\ref{DartmouthCollege}}
John~E.\ Carlstrom,\textsuperscript{\ref{UniversityofChicago},\ref{ArgonneNationalLaboratory}}
Julien Carron,\textsuperscript{\ref{UniversityofGeneva}}
Thomas Cecil,\textsuperscript{\ref{ArgonneNationalLaboratory}}
Anthony Challinor,\textsuperscript{\ref{UniversityofCambridge}}
Victor Chan,\textsuperscript{\ref{UniversityofToronto}}
Clarence~L.\ Chang,\textsuperscript{\ref{ArgonneNationalLaboratory},\ref{UniversityofChicago}}
Scott Chapman,\textsuperscript{\ref{UniversityofBritishColumbia}}
Eric Charles,\textsuperscript{\ref{SLAC}}
Eric Chauvin,\textsuperscript{\ref{EricChauvinConsultingEngineer}}
Cheng Cheng,\textsuperscript{\ref{UniversityofKwazulu-Natal}}
Grace Chesmore,\textsuperscript{\ref{UniversityofChicago}}
Kolen Cheung,\textsuperscript{\ref{UCBerkeley},\ref{LawrenceBerkeleyNationalLaboratory}}
Yuji Chinone,\textsuperscript{\ref{UniversityofTokyo}}
Jens Chluba,\textsuperscript{\ref{UniversityofManchester}}
Hsiao-Mei Sherry Cho,\textsuperscript{\ref{SLAC}}
Steve Choi,\textsuperscript{\ref{CornellUniversity}}
Justin Clancy,\textsuperscript{\ref{UniversityofMelbourne}}
Susan Clark,\textsuperscript{\ref{StanfordUniversity},\ref{KIPAC}}
Asantha Cooray,\textsuperscript{\ref{UCIrvine}}
Gabriele Coppi,\textsuperscript{\ref{UniversityofMilano-Bicocca}}
John Corlett,\textsuperscript{\ref{LawrenceBerkeleyNationalLaboratory}}
Will Coulton,\textsuperscript{\ref{FlatironInstitute}}
Thomas~M. Crawford,\textsuperscript{\ref{UniversityofChicago},\ref{KICP}}
Abigail Crites,\textsuperscript{\ref{CornellUniversity},\ref{Caltech}}
Ari Cukierman,\textsuperscript{\ref{SLAC},\ref{StanfordUniversity}}
Francis-Yan Cyr-Racine,\textsuperscript{\ref{UniversityofNewMexico}}
Wei-Ming Dai,\textsuperscript{\ref{UniversityofKwazulu-Natal}}
Cail Daley,\textsuperscript{\ref{UniversityofIllinoisatUrbana-Champaign}}
Eli Dart,\textsuperscript{\ref{LawrenceBerkeleyNationalLaboratory}}
Gregorg Daues,\textsuperscript{\ref{UniversityofIllinoisatUrbana-Champaign}}
Tijmen de Haan,\textsuperscript{\ref{KEK}}
Cosmin Deaconu,\textsuperscript{\ref{UniversityofChicago},\ref{KICP}}
Jacques Delabrouille,\textsuperscript{\ref{CentrePierreBinétruy}}
Greg Derylo,\textsuperscript{\ref{Fermilab}}
Mark Devlin,\textsuperscript{\ref{UniversityofPennsylvania}}
Eleonora Di Valentino,\textsuperscript{\ref{UniversityofSheffield}}
Marion Dierickx,\textsuperscript{\ref{HarvardUniversity}}
Brad Dober,\textsuperscript{\ref{NIST}}
Randy Doriese,\textsuperscript{\ref{NIST}}
Shannon Duff,\textsuperscript{\ref{NIST}}
Daniel Dutcher,\textsuperscript{\ref{PrincetonUniversity}}
Cora Dvorkin,\textsuperscript{\ref{HarvardUniversity}}
Rolando Dünner,\textsuperscript{\ref{PontificiaUniversidadCatólicadeChile}}
Tarraneh Eftekhari,\textsuperscript{\ref{NorthwesternUniversity}}
Joseph Eimer,\textsuperscript{\ref{JohnsHopkinsUniversity}}
Hamza El Bouhargani,\textsuperscript{\ref{LawrenceBerkeleyNationalLaboratory}}
Tucker Elleflot,\textsuperscript{\ref{LawrenceBerkeleyNationalLaboratory}}
Nick Emerson,\textsuperscript{\ref{UniversityofArizona}}
Josquin Errard,\textsuperscript{\ref{AstroParticleandCosmologyLaboratory}}
Thomas Essinger-Hileman,\textsuperscript{\ref{NASAGoddardSpaceFlightCenter}}
Giulio Fabbian,\textsuperscript{\ref{CardiffUniversity},\ref{FlatironInstitute}}
Valentina Fanfani,\textsuperscript{\ref{UniversityofMilano-Bicocca}}
Alessandro Fasano,\textsuperscript{\ref{LAMMarseille}}
Chang Feng,\textsuperscript{\ref{UniversityofIllinoisatUrbana-Champaign}}
Simone Ferraro,\textsuperscript{\ref{LawrenceBerkeleyNationalLaboratory}}
Jeffrey P.~Filippini,\textsuperscript{\ref{UniversityofIllinoisatUrbana-Champaign}}
Raphael Flauger,\textsuperscript{\ref{UCSanDiego}}
Brenna Flaugher,\textsuperscript{\ref{Fermilab}}
Aurelien A.~Fraisse,\textsuperscript{\ref{PrincetonUniversity}}
Josef Frisch,\textsuperscript{\ref{SLAC}}
Andrei Frolov,\textsuperscript{\ref{SimonFraserUniversity}}
Nicholas Galitzki,\textsuperscript{\ref{UCSanDiego}}
Patricio A.~Gallardo,\textsuperscript{\ref{UniversityofChicago}}
Silvia Galli,\textsuperscript{\ref{InstitutdAstrophysiquedeParis}}
Ken Ganga,\textsuperscript{\ref{AstroParticleandCosmologyLaboratory}}
Martina Gerbino,\textsuperscript{\ref{INFN}}
Christos Giannakopoulos,\textsuperscript{\ref{UniversityofCincinnati}}
Murdock Gilchriese,\textsuperscript{\ref{LawrenceBerkeleyNationalLaboratory}}
Vera Gluscevic,\textsuperscript{\ref{UniversityofSouthernCalifornia}}
Neil Goeckner-Wald,\textsuperscript{\ref{StanfordUniversity}}
David Goldfinger,\textsuperscript{\ref{HarvardUniversity}}
Daniel Green,\textsuperscript{\ref{UCSanDiego}}
Paul Grimes,\textsuperscript{\ref{CenterforAstrophysics|HarvardandSmithsonian}}
Daniel Grin,\textsuperscript{\ref{HaverfordCollege}}
Evan Grohs,\textsuperscript{\ref{NorthCarolinaStateUniversity}}
Riccardo Gualtieri,\textsuperscript{\ref{ArgonneNationalLaboratory}}
Vic Guarino,\textsuperscript{\ref{ArgonneNationalLaboratory}}
Jon E.~Gudmundsson,\textsuperscript{\ref{StockholmUniversity}}
Ian Gullett,\textsuperscript{\ref{CaseWesternReserveUniversity}}
Sam Guns,\textsuperscript{\ref{UCBerkeley}}
Salman Habib,\textsuperscript{\ref{ArgonneNationalLaboratory}}
Gunther Haller,\textsuperscript{\ref{SLAC}}
Mark Halpern,\textsuperscript{\ref{UniversityofBritishColumbia}}
Nils W.~Halverson,\textsuperscript{\ref{UniversityofColoradoBoulder}}
Shaul Hanany,\textsuperscript{\ref{UniversityofMinnesota}}
Emma Hand,\textsuperscript{\ref{UniversityofCincinnati}}
Kathleen Harrington,\textsuperscript{\ref{UniversityofChicago}}
Masaya Hasegawa,\textsuperscript{\ref{KEK}}
Matthew Hasselfield,\textsuperscript{\ref{FlatironInstitute}}
Masashi Hazumi,\textsuperscript{\ref{KEK}}
Katrin Heitmann,\textsuperscript{\ref{ArgonneNationalLaboratory}}
Shawn Henderson,\textsuperscript{\ref{SLAC}}
Brandon Hensley,\textsuperscript{\ref{PrincetonUniversity}}
Ryan Herbst,\textsuperscript{\ref{SLAC}}
Carlos Hervias-Caimapo,\textsuperscript{\ref{FloridaStateUniversity}}
J. Colin Hill,\textsuperscript{\ref{Columbia},\ref{FlatironInstitute}}
Richard Hills,\textsuperscript{\ref{Cavendish}}
Eric Hivon,\textsuperscript{\ref{InstitutdAstrophysiquedeParis},\ref{SorbonneUniversité}}
Ren\'{e}e Hlo\v{z}ek,\textsuperscript{\ref{UniversityofToronto},\ref{DunlapInstitute}}
Anna Ho,\textsuperscript{\ref{MillerInstitute},\ref{UCBerkeley}}
Gil Holder,\textsuperscript{\ref{UniversityofIllinoisatUrbana-Champaign}}
Matt Hollister,\textsuperscript{\ref{Fermilab}}
William Holzapfel,\textsuperscript{\ref{UCBerkeley}}
John Hood,\textsuperscript{\ref{UniversityofChicago}}
Selim Hotinli,\textsuperscript{\ref{JohnsHopkinsUniversity}}
Alec Hryciuk,\textsuperscript{\ref{UniversityofChicago}}
Johannes Hubmayr,\textsuperscript{\ref{NIST}}
Kevin~M.\ Huffenberger,\textsuperscript{\ref{FloridaStateUniversity}}
Howard Hui,\textsuperscript{\ref{Caltech}}
Roberto Ib\'{a}\~{n}ez,\textsuperscript{\ref{AssociatedUniversitiesIncorporated}}
Ayodeji Ibitoye,\textsuperscript{\ref{UniversityofKwazulu-Natal}}
Margaret Ikape,\textsuperscript{\ref{UniversityofToronto}}
Kent Irwin,\textsuperscript{\ref{StanfordUniversity}}
Cooper Jacobus,\textsuperscript{\ref{UCBerkeley}}
Oliver Jeong,\textsuperscript{\ref{UCBerkeley}}
Bradley R.~Johnson,\textsuperscript{\ref{UniversityofVirginia}}
Doug Johnstone,\textsuperscript{\ref{HerzbergAstronomyandAstrophysicsNRC},\ref{UniversityofVictoria}}
William C.~Jones,\textsuperscript{\ref{PrincetonUniversity}}
John Joseph,\textsuperscript{\ref{LawrenceBerkeleyNationalLaboratory}}
Baptiste Jost,\textsuperscript{\ref{AstroParticleandCosmologyLaboratory},\ref{CentrePierreBinétruy}}
Jae Hwan Kang,\textsuperscript{\ref{Caltech}}
Ari Kaplan,\textsuperscript{\ref{UCSantaBarbara}}
Kirit S.~Karkare,\textsuperscript{\ref{UniversityofChicago},\ref{Fermilab}}
Nobuhiko Katayama,\textsuperscript{\ref{KavliIPMU}}
Reijo Keskitalo,\textsuperscript{\ref{LawrenceBerkeleyNationalLaboratory},\ref{UCBerkeley}}
Cesiley King,\textsuperscript{\ref{WashingtonUniversitySt.Louis}}
Theodore Kisner,\textsuperscript{\ref{LawrenceBerkeleyNationalLaboratory},\ref{UCBerkeley}}
Matthias Klein,\textsuperscript{\ref{LMUMunich}}
Lloyd Knox,\textsuperscript{\ref{UCDavis}}
Brian J.~Koopman,\textsuperscript{\ref{YaleUniversity}}
Arthur Kosowsky,\textsuperscript{\ref{UniversityofPittsburgh}}
John Kovac,\textsuperscript{\ref{HarvardUniversity},\ref{CenterforAstrophysics|HarvardandSmithsonian}}
Ely D.~Kovetz,\textsuperscript{\ref{BenGurionUniversity}}
Alex Krolewski,\textsuperscript{\ref{LawrenceBerkeleyNationalLaboratory}}
Donna Kubik,\textsuperscript{\ref{Fermilab}}
Steve Kuhlmann,\textsuperscript{\ref{ArgonneNationalLaboratory}}
Chao-Lin Kuo,\textsuperscript{\ref{StanfordUniversity},\ref{SLAC}}
Akito Kusaka,\textsuperscript{\ref{LawrenceBerkeleyNationalLaboratory},\ref{UniversityofTokyo}}
Anne L\"ahteenm\"aki,\textsuperscript{\ref{AaltoUniversity}}
Kenny Lau,\textsuperscript{\ref{UniversityofMinnesota}}
Charles R.~Lawrence,\textsuperscript{\ref{JPL}}
Adrian T.~Lee,\textsuperscript{\ref{UCBerkeley},\ref{LawrenceBerkeleyNationalLaboratory}}
Louis Legrand,\textsuperscript{\ref{UniversityofGeneva}}
Matthaeus Leitner,\textsuperscript{\ref{LawrenceBerkeleyNationalLaboratory}}
Clément Leloup,\textsuperscript{\ref{AstroParticleandCosmologyLaboratory},\ref{CentrePierreBinétruy}}
Antony Lewis,\textsuperscript{\ref{UniversityofSussex}}
Dale Li,\textsuperscript{\ref{SLAC}}
Eric Linder,\textsuperscript{\ref{LawrenceBerkeleyNationalLaboratory},\ref{UCBerkeley}}
Ioannis Liodakis,\textsuperscript{\ref{UniversityofTurku}}
Jia Liu,\textsuperscript{\ref{KavliIPMU}}
Kevin Long,\textsuperscript{\ref{LawrenceBerkeleyNationalLaboratory}}
Thibaut Louis,\textsuperscript{\ref{IJCLab}}
Marilena Loverde,\textsuperscript{\ref{UniversityofWashington}}
Lindsay Lowry,\textsuperscript{\ref{UCBerkeley}}
Chunyu Lu,\textsuperscript{\ref{UniversityofIllinoisatUrbana-Champaign}}
Phil Lubin,\textsuperscript{\ref{UCSantaBarbara}}
Yin-Zhe Ma,\textsuperscript{\ref{UniversityofKwazulu-Natal}}
Thomas Maccarone,\textsuperscript{\ref{TexasTechUniversity}}
Mathew S.~Madhavacheril,\textsuperscript{\ref{PerimeterInstitute}}
Felipe Maldonado,\textsuperscript{\ref{FloridaStateUniversity}}
Adam Mantz,\textsuperscript{\ref{StanfordUniversity}}
Gabriela Marques,\textsuperscript{\ref{FloridaStateUniversity}}
Frederick Matsuda,\textsuperscript{\ref{ISASJAXA}}
Philip Mauskopf,\textsuperscript{\ref{ArizonaStateUniversity}}
Jared May,\textsuperscript{\ref{WashingtonUniversitySt.Louis}}
Heather McCarrick,\textsuperscript{\ref{PrincetonUniversity}}
Ken McCracken,\textsuperscript{\ref{CenterforAstrophysics|HarvardandSmithsonian}}
Jeffrey McMahon,\textsuperscript{\ref{UniversityofChicago},\ref{Fermilab}}
P.~Daniel Meerburg,\textsuperscript{\ref{UniversityofGroningen}}
Jean-Baptiste Melin,\textsuperscript{\ref{CEASaclay}}
Felipe Menanteau,\textsuperscript{\ref{UniversityofIllinoisatUrbana-Champaign}}
Joel Meyers,\textsuperscript{\ref{SouthernMethodistUniversity}}
Marius Millea,\textsuperscript{\ref{UCBerkeley}}
Vivian Miranda,\textsuperscript{\ref{StonyBrookUniversity}}
Don Mitchell,\textsuperscript{\ref{Fermilab}}
Joseph Mohr,\textsuperscript{\ref{LMUMunich}}
Lorenzo Moncelsi,\textsuperscript{\ref{Caltech}}
Maria Elena Monzani,\textsuperscript{\ref{SLAC},\ref{StanfordUniversity}}
Magdy Moshed,\textsuperscript{\ref{AstroParticleandCosmologyLaboratory},\ref{CentrePierreBinétruy}}
Tony Mroczkowski,\textsuperscript{\ref{EuropeanSouthernObservatory}}
Suvodip Mukherjee,\textsuperscript{\ref{PerimeterInstitute},\ref{UniversityofAmsterdam}}
Moritz Münchmeyer,\textsuperscript{\ref{UniversityofWisconsinMadison}}
Daisuke Nagai,\textsuperscript{\ref{YaleUniversity}}
Chandan Nagarajappa,\textsuperscript{\ref{UniversityofKwazulu-Natal}}
Johanna Nagy,\textsuperscript{\ref{WashingtonUniversitySt.Louis}}
Toshiya Namikawa,\textsuperscript{\ref{KavliIPMU}}
Federico Nati,\textsuperscript{\ref{UniversityofMilano-Bicocca}}
Tyler Natoli,\textsuperscript{\ref{UniversityofChicago},\ref{KICP}}
Simran  Nerval,\textsuperscript{\ref{UniversityofToronto}}
Laura Newburgh,\textsuperscript{\ref{YaleUniversity}}
Hogan Nguyen,\textsuperscript{\ref{Fermilab}}
Erik Nichols,\textsuperscript{\ref{UniversityofChicago},\ref{RSS}}
Andrina Nicola,\textsuperscript{\ref{PrincetonUniversity}}
Michael D.~Niemack,\textsuperscript{\ref{CornellUniversity}}
Brian Nord,\textsuperscript{\ref{Fermilab}}
Tim Norton,\textsuperscript{\ref{CenterforAstrophysics|HarvardandSmithsonian}}
Valentine Novosad,\textsuperscript{\ref{ArgonneNationalLaboratory}}
Roger O'Brient,\textsuperscript{\ref{JPL},\ref{Caltech}}
Yuuki Omori,\textsuperscript{\ref{UniversityofChicago}}
Giorgio  Orlando ,\textsuperscript{\ref{UniversityofGroningen}}
Benjamin  Osherson,\textsuperscript{\ref{UniversityofIllinoisatUrbana-Champaign}}
Rachel  Osten,\textsuperscript{\ref{SpaceTelescopeScienceInstitute},\ref{JohnsHopkinsUniversity}}
Stephen Padin,\textsuperscript{\ref{Caltech}}
Scott Paine,\textsuperscript{\ref{CenterforAstrophysics|HarvardandSmithsonian}}
Bruce Partridge,\textsuperscript{\ref{HaverfordCollege}}
Sanjaykumar Patil,\textsuperscript{\ref{UniversityofSouthernCalifornia}}
Don Petravick,\textsuperscript{\ref{UniversityofIllinoisatUrbana-Champaign}}
Matthew Petroff,\textsuperscript{\ref{HarvardUniversity}}
Elena Pierpaoli,\textsuperscript{\ref{UniversityofSouthernCalifornia}}
Mauricio Pilleux,\textsuperscript{\ref{EONSSpA}}
Levon Pogosian,\textsuperscript{\ref{SimonFraserUniversity}}
Karthik Prabhu,\textsuperscript{\ref{UCDavis}}
Clement Pryke,\textsuperscript{\ref{UniversityofMinnesota}}
Giuseppe Puglisi,\textsuperscript{\ref{UniversityofRomeTorVergata},\ref{LawrenceBerkeleyNationalLaboratory}}
Benjamin Racine,\textsuperscript{\ref{Aix-MarseilleUniversity}}
Srinivasan Raghunathan,\textsuperscript{\ref{UniversityofIllinoisatUrbana-Champaign}}
Alexandra Rahlin,\textsuperscript{\ref{Fermilab},\ref{UniversityofChicago}}
Marco Raveri,\textsuperscript{\ref{UniversityofPennsylvania}}
Ben Reese,\textsuperscript{\ref{SLAC}}
Christian L.~Reichardt,\textsuperscript{\ref{UniversityofMelbourne}}
Mathieu Remazeilles,\textsuperscript{\ref{InstitutodeFísicadeCantabria}}
Arianna Rizzieri,\textsuperscript{\ref{AstroParticleandCosmologyLaboratory},\ref{CentrePierreBinétruy}}
Graca Rocha,\textsuperscript{\ref{JPL},\ref{Caltech}}
Natalie A.~Roe,\textsuperscript{\ref{LawrenceBerkeleyNationalLaboratory}}
Kaja Rotermund,\textsuperscript{\ref{LawrenceBerkeleyNationalLaboratory}}
Anirban Roy,\textsuperscript{\ref{CornellUniversity}}
John E.~Ruhl,\textsuperscript{\ref{CaseWesternReserveUniversity}}
Joe Saba,\textsuperscript{\ref{LawrenceBerkeleyNationalLaboratory}}
Noah Sailer,\textsuperscript{\ref{UCBerkeley}}
Maria Salatino,\textsuperscript{\ref{StanfordUniversity}}
Benjamin Saliwanchik,\textsuperscript{\ref{BrookhavenNationalLaboratory}}
Leonid Sapozhnikov,\textsuperscript{\ref{SLAC}}
Mayuri Sathyanarayana Rao,\textsuperscript{\ref{RamanResearchInstitute}}
Lauren Saunders,\textsuperscript{\ref{YaleUniversity}}
Emmanuel Schaan,\textsuperscript{\ref{LawrenceBerkeleyNationalLaboratory}}
Alessandro Schillaci,\textsuperscript{\ref{Caltech}}
Benjamin Schmitt,\textsuperscript{\ref{HarvardUniversity}}
Douglas Scott,\textsuperscript{\ref{UniversityofBritishColumbia}}
Neelima Sehgal,\textsuperscript{\ref{StonyBrookUniversity}}
Sarah Shandera,\textsuperscript{\ref{PennsylvaniaStateUniversity}}
Blake~D.\ Sherwin,\textsuperscript{\ref{UniversityofCambridge}}
Erik Shirokoff,\textsuperscript{\ref{UniversityofChicago}}
Corwin  Shiu,\textsuperscript{\ref{PrincetonUniversity}}
Sara M.~Simon,\textsuperscript{\ref{Fermilab}}
Baibhav Singari,\textsuperscript{\ref{UniversityofMinnesota}}
An\v{z}e Slosar,\textsuperscript{\ref{BrookhavenNationalLaboratory}}
David Spergel,\textsuperscript{\ref{PrincetonUniversity}}
Tyler~St.\ Germaine,\textsuperscript{\ref{HarvardUniversity}}
Suzanne T.~Staggs,\textsuperscript{\ref{PrincetonUniversity}}
Antony A.~Stark,\textsuperscript{\ref{CenterforAstrophysics|HarvardandSmithsonian}}
Glenn D.~Starkman,\textsuperscript{\ref{CaseWesternReserveUniversity}}
Bryan Steinbach,\textsuperscript{\ref{Caltech}}
Radek Stompor,\textsuperscript{\ref{CentrePierreBinétruy},\ref{AstroParticleandCosmologyLaboratory}}
Chris Stoughton,\textsuperscript{\ref{Fermilab}}
Aritoki Suzuki,\textsuperscript{\ref{LawrenceBerkeleyNationalLaboratory}}
Osamu Tajima,\textsuperscript{\ref{KyotoUniversity}}
Chris Tandoi,\textsuperscript{\ref{UniversityofIllinoisatUrbana-Champaign}}
Grant P.~Teply,\textsuperscript{\ref{UCSanDiego}}
Gregg Thayer,\textsuperscript{\ref{SLAC},\ref{StanfordUniversity}}
Keith Thompson,\textsuperscript{\ref{StanfordUniversity}}
Ben Thorne,\textsuperscript{\ref{UCDavis}}
Peter Timbie,\textsuperscript{\ref{UniversityofWisconsinMadison}}
Maurizio Tomasi,\textsuperscript{\ref{UniversitadegliStudidiMilan},\ref{INFN}}
Cynthia Trendafilova,\textsuperscript{\ref{SouthernMethodistUniversity}}
Matthieu Tristram,\textsuperscript{\ref{IJCLab}}
Carole Tucker,\textsuperscript{\ref{CardiffUniversity}}
Gregory Tucker,\textsuperscript{\ref{BrownUniversity}}
Caterina Umilt\`{a},\textsuperscript{\ref{UniversityofIllinoisatUrbana-Champaign}}
Alexander van Engelen,\textsuperscript{\ref{ArizonaStateUniversity}}
Joshiwa van Marrewijk,\textsuperscript{\ref{EuropeanSouthernObservatory}}
Eve M.~Vavagiakis,\textsuperscript{\ref{CornellUniversity}}
Clara Vergès,\textsuperscript{\ref{HarvardUniversity}}
Joaquin D.~Vieira,\textsuperscript{\ref{UniversityofIllinoisatUrbana-Champaign}}
Abigail G.~Vieregg,\textsuperscript{\ref{UniversityofChicago}}
Kasey Wagoner,\textsuperscript{\ref{PrincetonUniversity}}
Benjamin Wallisch,\textsuperscript{\ref{UCSanDiego},\ref{InstituteforAdvancedStudy}}
Gensheng Wang,\textsuperscript{\ref{ArgonneNationalLaboratory}}
Guo-Jian Wang,\textsuperscript{\ref{UniversityofKwazulu-Natal}}
Scott Watson,\textsuperscript{\ref{SyracuseUniversity}}
Duncan Watts,\textsuperscript{\ref{UniversityofOslo}}
Chris Weaver,\textsuperscript{\ref{MichiganStateUniversity}}
Lukas Wenzl,\textsuperscript{\ref{CornellUniversity}}
Ben Westbrook,\textsuperscript{\ref{UCBerkeley}}
Martin White,\textsuperscript{\ref{LawrenceBerkeleyNationalLaboratory},\ref{UCBerkeley}}
Nathan Whitehorn,\textsuperscript{\ref{MichiganStateUniversity}}
Andrew Wiedlea,\textsuperscript{\ref{LawrenceBerkeleyNationalLaboratory}}
Paul Williams,\textsuperscript{\ref{LawrenceBerkeleyNationalLaboratory}}
Robert Wilson,\textsuperscript{\ref{SmithsonianAstrophysicalObservatory},\ref{BellLaboratories}}
Harrison Winch,\textsuperscript{\ref{UniversityofToronto}}
Edward J.\ Wollack,\textsuperscript{\ref{NASAGoddardSpaceFlightCenter}}
W.~L.~Kimmy Wu,\textsuperscript{\ref{SLAC},\ref{KIPAC}}
Zhilei Xu,\textsuperscript{\ref{MIT}}
Volodymyr G.~Yefremenko,\textsuperscript{\ref{ArgonneNationalLaboratory}}
Cyndia Yu,\textsuperscript{\ref{StanfordUniversity},\ref{SLAC}}
David Zegeye,\textsuperscript{\ref{UniversityofChicago}}
Jeff Zivick,\textsuperscript{\ref{UniversityofChicago}}
Andrea Zonca\textsuperscript{\ref{UCSanDiego}}

\begin{multicols}{2}
\scriptsize
\setlength{\parskip}{2pt}
% This file is automatically generated by a script, so edits may be overwritten. 
% author.py --optin ../2022_full_list/CMB-S4_Membership_Members_20220313_851am.csv --rulesfile CMB-S4_rules.pkl

\noindent\textsuperscript{\ref{UCIrvine}}\UCI

\noindent\textsuperscript{\ref{UniversityofMelbourne}}\Melbourne

\noindent\textsuperscript{\ref{JohnsHopkinsUniversity}}\JHU

\noindent\textsuperscript{\ref{UniversityofIllinoisatUrbana-Champaign}}\UrbanaC

\noindent\textsuperscript{\ref{SLAC}}\SLAC

\noindent\textsuperscript{\ref{ClemsonUniversity}}\Clemson

\noindent\textsuperscript{\ref{StanfordUniversity}}\Stanford

\noindent\textsuperscript{\ref{OxfordUniversity}}\Oxford

\noindent\textsuperscript{\ref{UCBerkeley}}\UCB

\noindent\textsuperscript{\ref{LawrenceBerkeleyNationalLaboratory}}\LBL

\noindent\textsuperscript{\ref{RiceUniversity}}\Rice

\noindent\textsuperscript{\ref{UniversityofBritishColumbia}}\UBC

\noindent\textsuperscript{\ref{Fermilab}}\FNAL

\noindent\textsuperscript{\ref{UCSanDiego}}\UCSD

\noindent\textsuperscript{\ref{CenterforAstrophysics|HarvardandSmithsonian}}\CfA

\noindent\textsuperscript{\ref{YaleUniversity}}\Yale

\noindent\textsuperscript{\ref{SISSA}}\SISSA

\noindent\textsuperscript{\ref{IFPU}}\IFPU

\noindent\textsuperscript{\ref{HarvardUniversity}}\HarvardPhys

\noindent\textsuperscript{\ref{UniversityofNewMexico}}\UNM

\noindent\textsuperscript{\ref{ArgonneNationalLaboratory}}\ANL

\noindent\textsuperscript{\ref{CardiffUniversity}}\Cardiff

\noindent\textsuperscript{\ref{AstroParticleandCosmologyLaboratory}}\APC

\noindent\textsuperscript{\ref{Caltech}}\Caltech

\noindent\textsuperscript{\ref{CornellUniversity}}\Cornell

\noindent\textsuperscript{\ref{NIST}}\NIST

\noindent\textsuperscript{\ref{UniversityofGeneva}}\Unige

\noindent\textsuperscript{\ref{InstitutdAstrophysiquedeParis}}\IAP

\noindent\textsuperscript{\ref{KICP}}\KICP

\noindent\textsuperscript{\ref{UniversityofChicago}}\UChicago

\noindent\textsuperscript{\ref{LAMMarseille}}\LAM

\noindent\textsuperscript{\ref{CentrePierreBinétruy}}\CPB

\noindent\textsuperscript{\ref{UniversityofCincinnati}}\Cincinnati

\noindent\textsuperscript{\ref{AssociatedUniversitiesIncorporated}}\AUI

\noindent\textsuperscript{\ref{JPL}}\JPL

\noindent\textsuperscript{\ref{LMUMunich}}\LMU

\noindent\textsuperscript{\ref{UniversityofTexasatAustin}}\UT

\noindent\textsuperscript{\ref{CITA}}\CITA

\noindent\textsuperscript{\ref{UniversitadiFerrara}}\UFerrara

\noindent\textsuperscript{\ref{INFN}}\INFN

\noindent\textsuperscript{\ref{UniversityofManchester}}\UoM

\noindent\textsuperscript{\ref{ArizonaStateUniversity}}\ASU

\noindent\textsuperscript{\ref{DartmouthCollege}}\Dartmouth

\noindent\textsuperscript{\ref{UniversityofCambridge}}\damtp

\noindent\textsuperscript{\ref{UniversityofToronto}}\daa

\noindent\textsuperscript{\ref{EricChauvinConsultingEngineer}}\EC

\noindent\textsuperscript{\ref{UniversityofKwazulu-Natal}}\KwaZuluNatal

\noindent\textsuperscript{\ref{UniversityofTokyo}}\KPMU

\noindent\textsuperscript{\ref{KIPAC}}\KIPAC

\noindent\textsuperscript{\ref{UniversityofMilano-Bicocca}}\MilanoBicocca

\noindent\textsuperscript{\ref{FlatironInstitute}}\CCA

\noindent\textsuperscript{\ref{KEK}}\KEK

\noindent\textsuperscript{\ref{UniversityofPennsylvania}}\UPenn

\noindent\textsuperscript{\ref{UniversityofSheffield}}\Sheffield

\noindent\textsuperscript{\ref{PrincetonUniversity}}\Princeton

\noindent\textsuperscript{\ref{PontificiaUniversidadCatólicadeChile}}\PU

\noindent\textsuperscript{\ref{NorthwesternUniversity}}\NWU

\noindent\textsuperscript{\ref{UniversityofArizona}}\UAS

\noindent\textsuperscript{\ref{NASAGoddardSpaceFlightCenter}}\GSFC

\noindent\textsuperscript{\ref{SimonFraserUniversity}}\SimonFraser

\noindent\textsuperscript{\ref{UniversityofSouthernCalifornia}}\SoCal

\noindent\textsuperscript{\ref{HaverfordCollege}}\Haverford

\noindent\textsuperscript{\ref{NorthCarolinaStateUniversity}}\NCSU

\noindent\textsuperscript{\ref{StockholmUniversity}}\OskarKlein

\noindent\textsuperscript{\ref{CaseWesternReserveUniversity}}\CWRU

\noindent\textsuperscript{\ref{UniversityofColoradoBoulder}}\CUBoulder

\noindent\textsuperscript{\ref{UniversityofMinnesota}}\UMN

\noindent\textsuperscript{\ref{FloridaStateUniversity}}\FSU

\noindent\textsuperscript{\ref{Columbia}}\Columbia

\noindent\textsuperscript{\ref{Cavendish}}\Cavendish

\noindent\textsuperscript{\ref{SorbonneUniversité}}\NPNHE

\noindent\textsuperscript{\ref{DunlapInstitute}}\dunlap

\noindent\textsuperscript{\ref{MillerInstitute}}\Miller

\noindent\textsuperscript{\ref{UniversityofVirginia}}\UVA

\noindent\textsuperscript{\ref{HerzbergAstronomyandAstrophysicsNRC}}\NRC

\noindent\textsuperscript{\ref{UniversityofVictoria}}\UVic

\noindent\textsuperscript{\ref{UCSantaBarbara}}\UCSB

\noindent\textsuperscript{\ref{KavliIPMU}}\IPMU

\noindent\textsuperscript{\ref{WashingtonUniversitySt.Louis}}\WUSL

\noindent\textsuperscript{\ref{UCDavis}}\UCD

\noindent\textsuperscript{\ref{UniversityofPittsburgh}}\Pitt

\noindent\textsuperscript{\ref{BenGurionUniversity}}\BenGurion

\noindent\textsuperscript{\ref{AaltoUniversity}}\Aalto

\noindent\textsuperscript{\ref{UniversityofSussex}}\SussexAstronomy

\noindent\textsuperscript{\ref{UniversityofTurku}}\Turku

\noindent\textsuperscript{\ref{IJCLab}}\ICJLab

\noindent\textsuperscript{\ref{UniversityofWashington}}\UW

\noindent\textsuperscript{\ref{TexasTechUniversity}}\TTU

\noindent\textsuperscript{\ref{PerimeterInstitute}}\PI

\noindent\textsuperscript{\ref{ISASJAXA}}\JAXA

\noindent\textsuperscript{\ref{UniversityofGroningen}}\VSI

\noindent\textsuperscript{\ref{CEASaclay}}\CEADAP

\noindent\textsuperscript{\ref{SouthernMethodistUniversity}}\SMU

\noindent\textsuperscript{\ref{StonyBrookUniversity}}\StonyBrook

\noindent\textsuperscript{\ref{EuropeanSouthernObservatory}}\ESO

\noindent\textsuperscript{\ref{UniversityofAmsterdam}}\AmsterdamAstro

\noindent\textsuperscript{\ref{UniversityofWisconsinMadison}}\UWMadison

\noindent\textsuperscript{\ref{RSS}}\RSS

\noindent\textsuperscript{\ref{SpaceTelescopeScienceInstitute}}\STSCI

\noindent\textsuperscript{\ref{EONSSpA}}\EONS

\noindent\textsuperscript{\ref{UniversityofRomeTorVergata}}\Rome

\noindent\textsuperscript{\ref{Aix-MarseilleUniversity}}\CPPM

\noindent\textsuperscript{\ref{InstitutodeFísicadeCantabria}}\Cantabria

\noindent\textsuperscript{\ref{BrookhavenNationalLaboratory}}\BNL

\noindent\textsuperscript{\ref{RamanResearchInstitute}}\RRI

\noindent\textsuperscript{\ref{PennsylvaniaStateUniversity}}\PSU

\noindent\textsuperscript{\ref{KyotoUniversity}}\Kyoto

\noindent\textsuperscript{\ref{UniversitadegliStudidiMilan}}\UNIMI

\noindent\textsuperscript{\ref{BrownUniversity}}\Brown

\noindent\textsuperscript{\ref{InstituteforAdvancedStudy}}\IAS

\noindent\textsuperscript{\ref{SyracuseUniversity}}\Syracuse

\noindent\textsuperscript{\ref{UniversityofOslo}}\Oslo

\noindent\textsuperscript{\ref{MichiganStateUniversity}}\MSU

\noindent\textsuperscript{\ref{SmithsonianAstrophysicalObservatory}}\SAO

\noindent\textsuperscript{\ref{BellLaboratories}}\Bell

\noindent\textsuperscript{\ref{MIT}}\MIT

\end{multicols}

\def\as#1{[{\bf AS:} {\it #1}] }

\eject
\pagenumbering{arabic} 
\setcounter{page}{1}

\section*{CMB-S4 Overview and Context}

The cosmic microwave background `Stage 4'  CMB-S4  project is 
the next generation ground-based cosmic microwave background experiment, 
designed to cross critical thresholds in our understanding of the origin and evolution of the Universe, from the highest energies at the dawn of time through the growth of structure to the present day. The CMB-S4 science case is spectacular: the search for primordial gravitational waves as predicted from inflation and the imprint of relic particles including neutrinos,
unique 
insights into dark energy and tests of gravity on large scales, elucidating the role of baryonic feedback on galaxy formation and evolution, opening up a window on the transient Universe at millimeter wavelengths, and even the exploration of the outer Solar System. 
The CMB-S4 sensitivity to primordial gravitational waves will probe physics at the highest energy
scales and cross a major theoretically motivated threshold in constraints on inflation. 
The CMB-S4 search for new light relic particles will shed light on the early Universe 10,000 times farther
back than current experiments can reach.
Finally, the CMB-S4 Legacy Survey, covering 60\% of the sky with
unprecedented sensitivity and angular resolution from centimeter- to millimeter-wave observing bands,
will have a profound and lasting impact on Astronomy and Astrophysics and provide a powerful complement to surveys at other wavelengths, such as the Vera Rubin Observatory (VRO) Legacy Survey of Space and Time and those conducted by the Nancy Grace Roman Space Telescope, Euclid, and others yet to be imagined.
We emphasize that these critical thresholds cannot be reached without the level of community and 
agency investment and commitment required by CMB-S4. In particular, the CMB-S4 science goals
are out of the reach of any projected precursor experiment by a significant margin.

The formal CMB-S4 Collaboration was established in 2018 with the ratification of the bylaws and election of the various officers including the collaboration Executive Team and Governing Board. As of March 2022 the Collaboration has 320 members, 76 of whom hold positions within the organizational structure. These members represent 114 institutions in 19 countries on 6 continents, including 27 US states.

CMB-S4 is a joint NSF and DOE project, with the construction phase expected to be funded as an NSF MREFC project and a DOE HEP MIE project. DOE Critical Decision 0 (CD-0) was approved in July 2019.  An integrated project office has been constituted and tasked with advancing the CMB-S4 project in the NSF MREFC Preliminary Design Phase and toward DOE Critical Decision CD-1.
Support for the Integrated Project Office is being provided in part by DOE through their lead lab LBNL, and by NSF through an MSRI-R1 award to the University of Chicago, as the lead NSF institution. 

CMB-S4 was recommended by the 2014 Particle Physics Project Prioritization Panel (P5) report {\it Building for Discovery: Strategic Plan for U.S. Particle Physics in the Global Context\/} and by the 2015 National Academies report {\it A Strategic Vision for NSF Investments in Antarctic and Southern Ocean Research}. The community further developed the science case in the 2016 {\it CMB-S4 Science Book\/} \cite{Abazajian:2016yjj} and surveyed the status of the technology in the 2017 {\it CMB-S4 Technology Book} \cite{TechBookarXiv170602464A}.  This work formed the foundation for the joint NSF-DOE Concept Definition Task Force (CDT), a subpanel of the Astronomy and Astrophysics Advisory Committee (AAAC), a FACA committee advising DOE, NASA, and NSF. The CDT report was enthusiastically accepted by the AAAC in October 2017.  

Building on the CDT report, the CMB-S4 Collaboration and the pre-Project Development Group, composed of experienced project leaders drawn primarily from the national laboratories, produced the comprehensive document, {\it The CMB-S4 Science Case, Reference Design, and Project Plan} \cite{Abazajian:2019}, which we refer to here as the Decadal Survey Report (\dsr).  The \dsr\  has been updated to reflect the advanced design in the {\it The CMB-S4 Preliminary Baseline Design Report}, hereafter the \pbdr, which is scheduled to be posted soon. The material presented in this white paper has been extracted from these major overview reports, as well as many detailed project documents.  These reports and numerous other reports, collaboration bylaws, workshop and working group wiki pages, email lists, and much more may be found at the website \url{https://CMB-S4.org}. We also refer the reader to the dedicated Snowmass White Paper on the CMB experimental program~\cite{Snowmass2021:CMBexp}. 

In November 2021, CMB-S4 was strongly recommended (with no caveats) by the 2020~Decadal Survey report {\it Pathways to Discovery in Astronomy and Astrophysics for the 2020s.}  The Technical, Risk, and Cost Evaluation (TRACE) that the survey had performed by the Aerospace Corporation provided a cost estimate similar to ours and evaluated the project risk as medium-low.  The Decadal Survey recommendation reads ``The National Science Foundation and the Department of Energy should jointly pursue the design and implementation of the next generation ground-based cosmic microwave background experiment (CMB-S4).''

To achieve its transformational science goals, CMB-S4 requires an enormous increase in sensitivity over all current CMB experiments combined, and roughly an order-of-magnitude increase over any projected precursor experiment. A significant and unique feature of CMB-S4 from the outset has been the use of multiple sites, specifically combining the two best currently developed sites on Earth for millimeter-wave observing: the high Atacama Plateau in Chile and the geographical South Pole. The design of CMB-S4 exploits key features of these two sites, namely the ability to drill deep on a single small patch of the sky through an extraordinarily stable atmosphere from the South Pole, and the ability to survey up to 80\% of the sky from the exceptionally high and dry Atacama site. 

Current experimental efforts at these two sites are already being consolidated into two major precursor observatories to CMB-S4, the Simons Observatory (SO) and the South Pole Observatory (SPO), members of whose teams also make up a large part of the CMB-S4 collaboration. The timing of both of these observatories is well-aligned with CMB-S4, enabling them to act as valuable pathfinders for CMB-S4 by providing technical and scientific data that have informed and will continue to inform our design and operations. To this end, both have also provided Letters of Intent to share their technical and cost data with CMB-S4. Nonetheless, while both will make significant advances in key CMB science goals, they will still fall well short of the thresholds targeted by CMB-S4. For example, to match the sensitivity to primordial gravitational waves provided by the ultra-deep CMB-S4 survey, SPO would have to 
integrate for nearly 50 years; it would take SO a similar amount of time to match the sensitivity to light relics provided by the CMB-S4 deep and wide survey. 

From space, the Japanese LiteBIRD CMB satellite mission was selected by JAXA for launch late in the decade for a 3-year mission, concurrent with CMB-S4 operations. With its lower resolution but wider frequency coverage, LiteBIRD's science goals are distinct from but highly complementary to CMB-S4's, and we are already discussing the parameters of a possible Memorandum of Understanding to enable both experiments to enhance their reach using elements of the other's data.

In short, CMB-S4 will enable transformational science that cannot be achieved otherwise, the CMB-S4 concept has clear community and agency support, and the CMB-S4 collaboration and project are moving forward.

\section*{Key Science Goals and Objectives}
\label{sec:keyScience}

The rich and diverse set of CMB-S4 scientific goals are organized into four themes: 
\begin{enumerate}
\setlength{\itemsep}{0mm} 
\setlength{\parsep}{0mm}
\item \textit{primordial gravitational waves and inflation};
\item \textit{the dark Universe};
\item \textit{mapping matter in the cosmos};
\item \textit{the time-variable millimeter-wave sky}. 
\end{enumerate}
The first two science themes relate to fundamental physics, and are of particular interest to the particle physics community. The other two themes relate to the broader scientific opportunities made possible by a millimeter-wave survey of unprecedented depth and breadth. Here we briefly review the key high-level goals and refer the reader to the more detailed science case in the \dsr, \pbdr, and in the Decadal Survey and Snowmas2021 science white papers referenced.
 
\paragraph{Primordial gravitational waves and inflation.}

We have a historic opportunity to open up a window to the primordial Universe \cite{swp-shandera}. If the predictions of some of the leading models for the origin of the hot big bang are borne out, CMB-S4 will detect the signature of primordial gravitational waves in the polarization pattern of the CMB \cite{CMB-S4_pgw2022}. This detection would provide the first evidence for the quantization of gravity, reveal new physics at the energy scale of grand unified theories, and yield insight into the symmetries of nature.

The current leading scenario for the origin of structure in our Universe is cosmic inflation, a period of accelerated expansion prior to the hot big bang, as discussed in the dedicated Snowmass 2021 White Papers~\cite{Snowmass2021:Inflation,Snowmass2021:TheoryCosmo}, also see \cite{Inflation-NonG,Inflation-PowerS}. During this epoch, quantum fluctuations were imprinted on all spatial scales in the Universe. These fluctuations seeded the density perturbations that developed into all the structure in the Universe today. While 
there are still viable alternative models for the early history of the Universe, 
the simplest models of inflation are exceptionally successful in describing the data. 

Tantalizingly, the observed scale dependence of the amplitude of density perturbations has quantitative implications for the amplitude of primordial gravitational waves, commonly parameterized by $r$, the ratio of fluctuation power in gravitational waves to that in density perturbations. All inflation models that naturally explain the observed deviation from scale invariance and that also have a characteristic scale equal
to or larger than the Planck scale 
predict $r \gtrsim 0.001$. A well-motivated sub-class within this set of models
is detectable by CMB-S4 at 5$\sigma$.  
The observed departure from scale invariance is a potentially important clue that strongly motivates exploring down to $r = 10^{-3}$.  With an order of magnitude more detectors than precursor observations, and exquisite control of systematic errors, CMB-S4 will improve upon limits from pre-CMB-S4 observations by a factor of five to reach this target, allowing us to either detect primordial gravitational waves or 
rule out large classes of inflationary models and dramatically impact how we think about the theory.

\paragraph{The dark Universe.}
In the standard cosmological model, about 95\% of the energy density of the Universe is in dark matter and dark energy.  As discussed in the dedicated Snowmass 2021 White Papers~\cite{Snowmass2021:CMB-S4_DM,Snowmass2021:CosmoLabNeutrinos,Snowmass2021:Flagship_DE}, with CMB-S4 we can address numerous questions about these dark ingredients, such as: How is matter distributed on large scales? Does the dark matter have non-gravitational interactions with baryons? Are there additional unseen components beyond dark matter and dark energy? 

Light relic particles are one very well-motivated possibility for additional energy density, as additional light particles appear frequently and numerously in extensions to the standard model of particle physics \cite{swp-green,Snowmass2021:LightRelics}. For large regions of the unexplored parameter space in these models, the light particles are thermalized in the early Universe.
The Planck satellite has sensitivity to light particles that fell out of thermal equilibrium after the first 50 micro-seconds of the Universe. With CMB-S4 we can push back this frontier by over a factor of 10,000, to the first fractions of a nanosecond. 

The contribution of light relics to the energy density, often parameterized as the ``effective number of neutrino species,'' $\Neff$, leads to observable consequences in the CMB temperature and polarization anisotropy. Current data are only sensitive enough to detect additional relics that froze out after the quark-hadron transition, 
so CMB-S4's ability to probe times well before that transition is a major advance. Specifically CMB-S4 will constrain $\Delta \Neff < 0.06$ at 95\% C.L., achieving sensitivity to Weyl fermion and vector particles that froze out at temperatures a few hundred times higher than that of the QCD phase transition.

CMB-S4 will also enable a broader exploration of the dark Universe in combination with other probes, often significantly enhancing them by breaking their intrinsic degeneracies.  It will improve or detect various possibilities for the properties of dark matter beyond the simplest cold dark matter models \cite{swp-gluscevic,Snowmass2021:StaticProbes}. 
It will add to dark energy constraints through precision measurements of the primordial power spectrum, and
through precision measurements of the lensing convergence power spectrum, through the CMB-lensing-derived mass calibration of galaxy clusters \cite{swp-mantz}, and through CMB lensing tomography \cite{swp-slosar1}.

\paragraph{Mapping matter in the cosmos.}
Observations indicate there is 
roughly five times more dark matter than baryonic matter 
and that most of the  baryonic matter is in the form of hot ionized gas rather
than cold gas or stars.
CMB-S4 will be able to map out 
normal and dark 
matter separately by measuring the fluctuations in the
total mass density (using gravitational lensing) and the ionized gas density (using Compton scattering).

Observations of gravitational lensing of the CMB are key to many CMB-S4 science goals.
CMB-S4 lensing data will lead to a precise two-dimensional map of the total matter distribution.
The statistical properties of this mass map will provide important constraints on
dark energy \cite{swp-slosar1}, 
modified gravity \cite{swp-slosar1}, and the neutrino masses \cite{swp-dvorkin}.
When combined with CMB-S4-derived or external catalogs of galaxies or galaxy clusters, 
this mass map can be used to ``weigh'' the galaxy or cluster samples. 
With galaxies, this can be done in a redshift-dependent or tomographic manner  
 out to redshifts as high as $z \sim 5$,  
 making possible new precision tests of cosmology and gravity.
With robust CMB-lensing-based cluster masses
at high redshift, the abundance of galaxy clusters can be used as an additional
probe of dark energy and neutrino masses. 

Most of the baryons in the late Universe are believed to be in a diffuse ionized plasma 
that is difficult to observe \cite{swp-cicone,swp-oppenheimer,swp-wang}. 
CMB-S4 will measure the effect of Compton scattering by this gas 
(the Sunyaev-Zeldovich or SZ effects), both the spectral distortion from hot electrons
(thermal SZ or  tSZ) and a general redshift or blueshift of the scattered
photons due to coherent bulk flows along the line of sight (kinematic SZ or kSZ). 
The nature
of this scattering makes the SZ effects independent of redshift. 
With a deep and wide survey covering a large amount of volume and
an ultra-deep survey imaging lower-mass clusters, CMB-S4 will be an 
effective probe of 
the crucial regime of $z \gtrsim 2$, 
when galaxy clusters were vigorously accreting new hot gas while at the same time
forming the bulk of their stars \cite{swp-overzier}. 
The CMB-S4 catalog 
will contain an order of magnitude more clusters at $z > 2$ than will be discovered with Stage 3 CMB
experiments \cite{swp-mantz,swp-dannerbauer}. 
CMB-S4 will also measure the diffuse tSZ signal on the sky and make a 
temperature-weighted map of 
ionized gas 
that can be used to measure the average
thermal pressure profiles 
around 
galaxies and groups of galaxies. 
CMB-S4 will also make maps of the kSZ effect, which will be combined with data from 
other surveys to make maps of the projected electron density 
around samples of objects. 
Applications of these maps include 
measuring ionized gas as a function of radius, directly constraining
the impact of feedback from active galactic nuclei and supernovae on the 
intergalactic medium \cite{swp-battaglia} and
constraining theories of modified gravity using the bulk flow amplitude as a function of separation. 
Even without overlapping galaxy catalogs, the kSZ signal can be used to probe 
the epoch of reionization, in ways that are highly complementary to the 
measurements of the neutral gas that can be obtained with redshifted Ly-$\alpha$
and 21-cm studies \cite{swp-chang,swp-cooray,swp-hutter,swp-laplante}.

\paragraph{The time-variable millimeter-wave sky.}

There have been relatively few studies of the variable sky at millimeter wavelengths, with
systematic surveys using CMB data only very recently being undertaken \cite{swp-holder, Whitehorn2016, Naess:2020qbb, SPT-3G:2021vsz}).  
A deep, wide, millimeter-wave survey with time-domain capability will provide key insights
into transient or burst events, moving sources such as Solar-System objects, and variable
sources such as stars and active galactic nuclei (AGN).

Targeted follow-up observations of gamma-ray bursts, core-collapse supernovae, tidal disruption events,
classical novae, X-ray binaries, and stellar flares have found that there are many transient events
with measured fluxes that would make them detectable by CMB-S4.  A systematic survey of the mm-wave sky with a
cadence of a day or two over a large fraction of the sky, combined with
an ultra-deep daily survey of a few percent of the sky, would be an excellent
complement to other transient surveys, filling a gap between
radio and optical searches \cite{swp-holder}. Gamma-ray burst afterglows are particularly interesting targets as they peak at millimeter wavelengths and there is a possibility
of capturing mm-wave afterglows that have no corresponding gamma-ray trigger, either
from the geometry of relativistic beaming and/or from sources at 
very high redshift \cite{swp-holder}. Both are predicted theoretically, but have never been detected.

\commentout{
\begin{figure}
\begin{center}  
\includegraphics[width=6in]{Figures/transient_surveys.pdf}
\end{center}
\caption{Filled circles show 5$\sigma$ limiting magnitude for 
transient surveys with roughly daily cadence (ASAS-SN, Zwicky Transient Facility, Large Synoptic
Sky Survey, Australia SKA Pathfinder, all in black, CMB-S4 in red) over a large fraction of the
sky. Diagonal lines indicate constant $\nu f_\nu$, lines separated by factors of 100, green
shows a Neptune-mass planet at 700 AU, blue lines show SEDs corresponding to a quasar or blazar,
normalized to be representative in flux of the population measurable in daily CMB-S4 maps. Open
circles show coadded $5\sigma$ depth for past and near-future surveys that cover large fractions of the sky,
with CMB-S4 again shown in red.}
\label{fig:transients}
\end{figure}
}

\commentout{
Thermal emission from planets, 
dwarf planets, and a selection of asteroids has been measured at these wavelengths; 
since these
sources move across the sky they can be differentiated from the 
stationary extrasolar sky. 
CMB-S4 will provide a long well-sampled time baseline and a wavelength range that is 
well-suited for the detection of possible large objects in the outer Solar System.
These measurements will be highly complementary
to those using optical reflected light or thermal emission at infrared wavelengths.
}

CMB-S4 will play an active role in multi-messenger astronomy, providing a long
baseline with high-cadence sampling in both intensity and linear polarization over
a wide sky area.
For example, the IceCube event IC170922A is believed to be associated with a flaring gamma-ray state of the blazar TXS 0506+056. In December 2014, however, the same source appears to have had a neutrino luminosity at least 10 times larger with no associated gamma emission---and no data existed at other wavelengths.
Having high-cadence wide-field non-gamma-ray data will be critical to understand sources like this one.
Any similar source is likely to be included in 
CMB-S4's near-daily, high-signal-to-noise monitoring 
of the blazar population.
The wide-area nature of the survey will also make it straightforward
to search for gravitational wave sources, particularly for
sources that happen to be poorly localized and are challenging for other instruments.

\section*{Technical Overview}
\label{sec:techOverview}

The CMB-S4 Collaboration and Project have
developed a Preliminary Baseline Design that meets the measurement requirements and therefore can deliver the CMB-S4 science goals. 
The main components of the Preliminary Baseline Design are described in detail in the \pbdr\ and summarized here. The major components are as follows:

\begin{itemize}

\item An ultra-deep survey covering 3\% of the sky (more if a gravitational wave signal is detected) to be conducted over seven years
using: fourteen 0.55-m refractor small-aperture telescopes (SATs) at 155\,GHz and below and four 0.44-m SATs at 230/280\,GHz, 
with dichroic, horn-coupled superconducting transition-edge-sensor (TES) detectors in each SAT, measuring two of the eight targeted frequency bands between 
25 and 280\,GHz; and one 5-m class ``delensing" large-aperture telescope (LAT), 
equipped with detectors distributed over seven bands from 20 to 280\,GHz.
Measurements at degree angular scales and larger made using refractor telescopes with roughly 0.5-m apertures have been demonstrated to deliver high-fidelity, low-contamination polarization measurements at these scales.  The combination of the SATs with the 5-m LAT therefore provides low-resolution $B$-mode measurements with excellent control of systematic contamination, as well as the high-resolution measurements required for delensing.  The ultra-deep survey SATs and 5-m LAT are to be located at the South Pole to allow targeted observations of the single small-area field, with provisions to relocate a fraction of the SATs in Chile if, for example, a high level of $r$ is detected.

The total detector count for the 18 SATs is 
147,936,  
with the majority of the detectors allocated to the 85 to 155\,GHz bands.  
The total number of science-grade 150-mm detector wafers required for 18 SATs is 
216.
The delensing LAT will have a total TES detector count of
129,024,
with the majority of the detectors allocated to the 90 to 150\,GHz bands. 
The total number of science-grade 150-mm diameter detector wafers required for this single LAT is 85.  

\item A deep and wide survey covering approximately 60\% of the sky to be conducted over seven years using two 6-m 
LATs located in Chile, each equipped with 
137,996
TES detectors distributed over eight frequency bands spanning 25 to 280\,GHz. The total number of science-grade 150-mm diameter detector wafers required for these two LATs is 170.
\end{itemize}

In the context of their legacy value to the wider community, we refer to the deep/wide and ultra-deep high-resolution surveys together as the CMB-S4 Legacy Survey.
The total detector count for CMB-S4 is 
552,952, 
requiring 
471
science grade wafers. This is an enormous increase over the detector count of all Stage-3 experiments combined. Such a dramatic increase in scale is required to meet the CMB-S4 science goals. 

\section*{Technical Readiness}

The CMB-S4 reference design uses existing, well-demonstrated technology that has been developed and demonstrated by the CMB experimental groups over the last decade, scaled up to unprecedented levels. The design and implementation plan addresses the considerable technical challenges presented by the required scaling up of the instrumentation and by the scope and complexity of the data analysis and interpretation.  Features of the design and plan include: scaled-up superconducting TES detector arrays with well-understood and robust material properties and processing techniques; high-throughput mm-wave telescopes and optics with unprecedented precision and rejection of systematic contamination; full internal characterization of astronomical foreground emission; large cosmological simulations and improved theoretical modeling; and computational methods for extracting minute correlations in massive, multi-frequency data sets, which include noise and a host of known and unknown signals. 

A CMB-S4 Risk and Opportunity Management Plan describes the continuous risk and opportunity management process implemented by the project, consistent with DOE O413.3B, ``Project Management for the Acquisition of Capital Assets," and the NSF 21-107, ``Research Infrastructure Guide." The plan establishes the methods of assessing CMB-S4 project risk and opportunities for all subsystems as well as the system as a whole. The CMB-S4 risk register has 211 risks identified. There are 26 risks currently assessed as High. The project is working on mitigations to ensure that these risks are lowered to reasonable levels on a timescale consistent with the overall project schedule.

For example, a current identified High risk is meeting the scaled-up production throughput and testing timeline of the transition-edge-sensor detector arrays. This is a major focus of the R\&D program supported by the DOE. The Integrated Project Office formed a CMB-S4 Detector Fabrication Group (CDFG) in January 2020 to facilitate the collaboration between multiple fabrication sites and to develop a single detector fabrication plan and produce prototype detectors that satisfy CMB-S4 acceptance criteria. The CDFG is envisioned to be active throughout the duration of CMB-S4 detector fabrication efforts.

\section*{Organization, Partnerships, and Current Status}

CMB-S4 is both a scientific collaboration and a DOE/NSF project. While these are certainly tightly coupled, they do have different roles and responsibilities; the overall organization of CMB-S4 therefore decouples into the organization of the collaboration and the project.

The CMB-S4 Project has developed in tandem with the Collaboration, paced by the funding agencies. In 2019 DOE approved Critical Decision 0, identifying the project need, and NSF funded a project development proposal led by the University of Chicago. In 2020 DOE appointed Berkeley Lab as the DOE lead laboratory. Together Berkeley Lab and U.\ Chicago lead a single, integrated, joint project with 121 members, 53 of whom are also collaboration members. The project office has developed a wide range of project organization and documentation, including the Work Breakdown Structure, Risk and Opportunity Register, detailed Cost Book and Technically-Limited Schedule.  The top level WBS Structure is summarized in Table~\ref{tab:wbs_structure}.

\begin{table}[ht!]
\begin{center}
\caption[CMB-S4 WBS Structure]{CMB-S4 WBS Structure}
\label{tab:wbs_structure}
\small{
\begin{tabular}{| C{0.06\textwidth}| L{0.14\textwidth} | L{0.8\textwidth} |} 
\hline
WBS & WBS Title & WBS Description \cr
\hline
1.01 & Project Office &
Labor, travel, and materials necessary to plan, track, organize, manage, maintain communications, conduct reviews, and perform necessary safety, risk, and QA tasks during all phases of the project. Overall project Systems Engineering is a subsection of this WBS element. However, subsystem-related management and support activities for planning, estimating, tracking, and reporting as well as their specific EH\&S and QA tasks are included in each of the subsystems. \cr
\hline
1.03 & Detectors & 
Labor, materials, and equipment associated with the design, fabrication and testing of the detector wafers. R\&D activities to support development of a Conceptual Design pre CD-1 for DOE. \cr
\hline
1.04 & Readout &
Labor, materials, and equipment associated with the design, fabrication and testing of the detector readout system. R\&D activities to support development of a Conceptual Design pre CD-1 for DOE. \cr
\hline
1.05 & Module Assembly and Testing &
Labor, materials, and equipment associated with the design, parts fabrication, assembly and testing of the detector modules. R\&D activities to support development of a Conceptual Design pre CD-1 for DOE.\cr 
\hline
1.06 & Large Aperture Telescopes &
Labor, materials, and equipment associated with the design, prototyping, materials selection, construction and certification for the Large Aperture Telescope System. Integration and commissioning in North America.\cr
\hline
1.07 & Small Aperture Telescopes &
Labor, materials, and equipment associated with the design, prototyping, materials selection, construction and certification for the Small Aperture Telescope System. R\&D activities to support development of a Conceptual Design pre CD-1 for DOE.   Integration and commissioning in North America.\cr
\hline
1.08 & Data Acquisition and Control &
Labor, materials, and equipment associated with the design, construction, certification, and delivery of the control systems for the observatories and data acquisition. R\&D activities to support development of a Conceptual Design pre CD-1 for DOE. \cr
\hline
1.09 & Data Management &
Labor, materials, and equipment associated with the design, construction, certification, and delivery of the data management system. R\&D activities to support development of a Conceptual Design pre CD-1 for DOE. \cr
\hline
1.10 & Chile Infrastructure, Integration and Commissioning &
Labor, travel, and materials necessary to plan, track, manage, maintain communications, conduct reviews, and perform necessary safety monitoring on site including oversight of all shipping of CMB-S4 components to Chile and oversight of construction activities on site.  On-site Integration and Commissioning of the CMB-S4 telescopes and infrastructure in Chile.\cr
\hline
1.11 & South Pole Infrastructure, Integration and Commissioning & 
Labor, travel, and materials necessary to plan, track, manage, maintain communications, conduct reviews, and perform necessary safety monitoring on site including oversight of all shipping of CMB-S4 components to the South Pole and oversight of construction activities on site. On-site Integration and Commissioning of the CMB-S4 telescopes and infrastructure at the South Pole.\cr
\hline
\end{tabular}
}
\end{center}
\end{table}

\begin{figure}[htb]
\begin{center}
\includegraphics[width=0.98\textwidth]{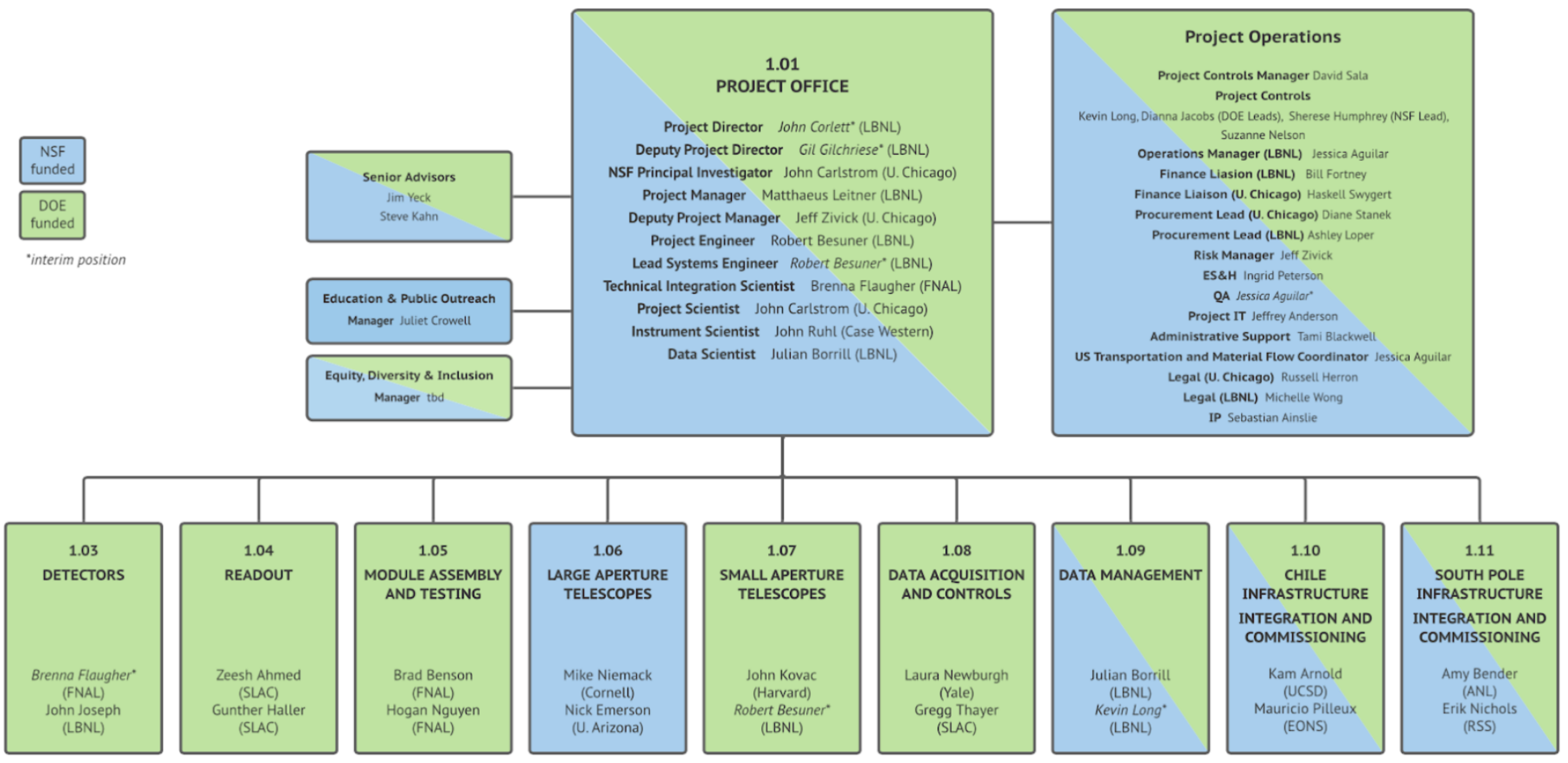}
\captionsetup{font=footnotesize}
\caption{Organizational Chart of the Integrated Project Office.  The figure includes the expected distribution of project scope by funding agency (NSF = blue, DOE = green).  We are actively pursuing partners who could make significant scope contributions in areas aligned with their expertise.}
\label{fig:prj_org}
\end{center}
\vskip -20pt
\end{figure}

The organizational chart of the Integrated Project Office is shown in Fig.~\ref{fig:prj_org}. A key feature of the organization is the role of collaboration members in the project office, in particular as leaders of the Level 2 systems.  The Level 2 managers are supported by engineering and project-management leaders.  The NSF/DOE scope distribution  promotes the engagement and participation of universities and national laboratories.  Graduate students, postdocs, professional technicians and engineers are expected to be involved in all aspects of the project.

The Integrated Project Office is responsible for forming partnerships with key stakeholder institutions, including DOE National Laboratories, universities, and potential collaborating projects such as the Simons Observatory, South Pole Observatory, and the CCAT-prime project.  Partnerships are also expected to include foreign institutions participating in the CMB-S4 Science Collaboration and contributing to the CMB-S4 Project.  

The CMB-S4 project is expected to include significant contributions from collaborating institutions supported by funding agencies other than NSF and DOE.  These ``in-kind'' contributions will be defined as deliverables to the project. Major contributions from partners will need to be negotiated and incorporated in the project design and Work Breakdown Structure within the next one to two years to avoid adding schedule and cost risk.

\section*{Schedule}

The CMB-S4 Project schedule estimate is developed from the work scope defined in the WBS and decomposing the work packages into detailed activities, estimating the duration of each activity, sequencing the activities in time, and establishing necessary predecessors and successors.  The data are captured in a formal Project Management Control System (PMCS), which encompasses the costs, schedules, work scope, and other activity attributes that define how the project will be executed.  Throughout the life of the project, control and analysis of the schedule requires the use of progress reporting, schedule change control systems, such as the use of project baseline change requests, earned value performance management, and variance analysis to determine if additional action is required to get the schedule back in line with the plan.

\begin{figure}[t!]
\begin{center}
\includegraphics[width=1.0\textwidth]{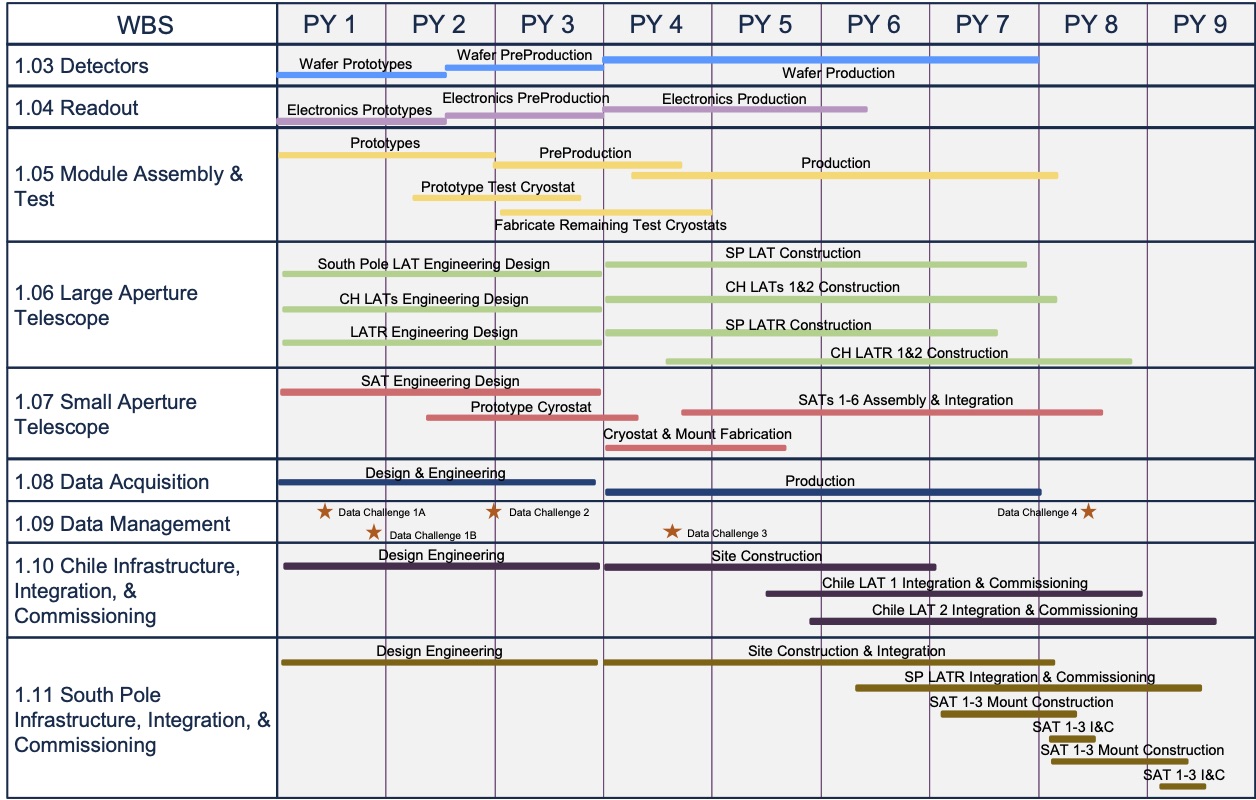}
\caption{Major activities by WBS for a technically-limited schedule for the CMB-S4 project. The timeline will depend on funding profile from DOE and NSF. To address the highest technical, cost, and schedule risks, the Detector, Readout, and Modules (DRM) R\&D, prototyping, and pre-production are prioritized to allow production to begin as early as possible. In parallel, telescope design and fabrication are synchronized with DRM delivery for assembly and test in the US before shipping to observation sites.  Integration and commissioning will be completed at the installed locations.}
\label{fig:wbs-milestones}
\end{center}
\end{figure}

The CMB-S4 project schedule has 8995 activities, 14,348 relationships, 21 Level 1/2, 110 Level 3, 179 Level 4, and 2179 Level 5 milestones. This preliminary schedule has been developed as a technically-limited schedule and will be updated once funding profile guidance from the DOE/NSF is available. Our current technically limited schedule is shown in Figure~\ref{fig:wbs-milestones}.

Seven years of operations are needed to achieve the CMB-S4 science goals.

\section*{Cost}

The CMB-S4 Project cost estimate has been developed from applying resource allocations to specific activities captured in the WBS, establishing budgets at the start of the project and to support budget status during the project.  The Cost Estimating Plan (CEP) focuses on the methodology to develop the activity-based estimates and facilitates the financial management and control of the Project as described in the Project Controls System Management Plan.

For task-based estimating, formal procedures are used to define tasks and assign them to the lowest-level elements of the WBS. The ensemble of tasks represents all the required resources, activities, and components of the entire project. Each of the tasks is included in the Integrated Project Schedule (IPS) and estimated by the teams using accepted techniques. The estimates are documented by a Basis of Estimate (BOE).
 
The cost estimates and schedule have been prepared by the IPO Project Managers, project controls specialists and the L2 Subsystem Control Account Managers and Scientists experienced in the specialized fields required to accomplish the CMB-S4 Project. Vendor estimates and quotations, engineering calculations, drawings, and other pertinent data, which are used to support the cost estimate, are collected and organized into a BOE. The details of the cost estimating process and assumptions are contained in the CMB-S4 Cost Estimating Plan.

The CMB-S4 point estimate project cost in January 2022 was \$636M for DOE Total Estimated Cost (TEC) funds plus NSF MREFC, fully loaded and escalated to the year of expenditure, and for a technically limited schedule without contingency. The Monte Carlo simulation contingency analysis of the CMB-S4 resource-loaded, risk-applied schedule yielded a budget contingency of \$266M. The cost are split roughly 50/50 between DOE and NSF scope.  DOE Other Project Costs (OPC) and NSF pre-MREFC funds total approximately \$61M. The IPO is continuing to improve the quality of the cost, schedule and funding profile of the CMB-S4 project in preparation for agency reviews. 

In-kind contributions from private and international partners are being pursued and, if realized, will reduce the total cost to NSF and DOE. 

\section*{Summary}  The science case for CMB-S4 remains as exciting as ever, as evidenced by DOE CD-0 ``Mission Need" approval and by the strong endorsement by the 2020 Decadal Survey of Astronomy and Astrophysics. In fact, the science case has become stronger as scientists spanning the particle physics and cosmology communities have explored the potential of the  transformational measurements CMB-S4 will make to achieve its threshold-crossing science goals.  The technology is well understood and field tested. The CMB-S4 Collaboration and Integrated Project Office are established and highly functioning. The project is well developed, has been extensively reviewed, and is on path to achieve DOE CD-1 and NSF PDR on the timescale of a year, depending on funding.

\eject

\markboth{\bibname}{}
\markright{\bibname}
\bibliography{cmbs4}

\end{document}